\documentclass{aa}   

\usepackage[comma,authoryear]{natbib}
\usepackage{graphicx}
\usepackage[varg]{txfonts} 
\usepackage{longtable}   
\usepackage{lscape}

\newcommand{\eg}{{e.g.}}
\newcommand{\ie}{{i.e.}}
\newcommand{\ms}{m.s$^{\rm -1}$}
\newcommand{\kms}{km.s$^{\rm -1}$}
\newcommand{\msy}{m.s$^{\rm -1}$.y$^{\rm -1}$}

\newcommand{\Mjup}{M$_{\rm Jup}$}
\newcommand{\ME}{M$_{\rm Earth}$}
\newcommand{\vsini}{$v\sin{i}$}
\newcommand{\Msun}{M$_{\sun}$}
\newcommand{\Rsun}{R$_{\sun}$}
\newcommand{\Mstar}{M$_{\rm \star}$}
\newcommand{\bv}{$B-V$}
\newcommand{\msini}{$m_{\rm p}\sin{i}$}
\newcommand{\rhk}{log$R'_{\rm HK}$}
\newcommand{\safir}{S{\small AFIR}}
\newcommand{\sophie}{S{\small OPHIE}}
\newcommand{\harps}{H{\small ARPS}}
\newcommand{\hipp}{H{\small IPPARCOS}}
\newcommand{\rosat}{R{\small OSAT}}
\newcommand{\pionier}{P{\small IONIER}}
\newcommand{\elodie}{E{\small LODIE}}

\begin{document}

   \title{Extrasolar planets and brown dwarfs around AF-type stars.
    \thanks{Based on observations collected at the European Southern Observatory, Chile, ESO 072.C-0636, 073.C-0733, 075.C-0689, 076.C-0279, 077.C-0295, 078.C-0209, 080.C-0664, 080.C-0712., 081.C-0774, 082.C-0412, 083.C-0794, 084.C-1039, 184.C-0815, 192.C-0224.}}

 \subtitle{IX. {\bf The HARPS Southern sample}}

   \author{
     S. Borgniet \inst{1}
     \and
     A.-M. Lagrange \inst{1}
      \and
     N. Meunier \inst{1}
     \and
     F. Galland \inst{1}
    }

   \institute{
Univ. Grenoble Alpes, IPAG, F-38000 Grenoble, France\\ 
CNRS, IPAG, F-38000 Grenoble, France\\
\email{simon.borgniet@univ-grenoble-alpes.fr}
    }

   \date{Received date / Accepted date}

   
   \abstract
   {Massive, Main-Sequence AF-type stars have so far remained unexplored in past RV surveys, due to their small number of spectral lines and their high rotational velocities that prevent the classic RV computation method.}
   {Our aim is to search for giant planets (GP) around massive AF Main-Sequence stars, to get first statistical information on their occurrence rate and to compare the results with evolved stars and lower-mass Main-Sequence stars.}
   {We used the \harps~spectrograph located on the 3.6m telescope at ESO La Silla Observatory to observe 108 AF Main-Sequence stars with \bv~in the range $-$0.04 to 0.58 and masses in the range 1.1-3.6~\Msun. We used our SAFIR software developed to compute the RV and other spectroscopic observables of these early-type stars. We characterized the detected companions as well as the intrinsic stellar variability. We computed the detection limits and used them as well as the detected companions to derive the first estimate of the close-in brown dwarf and giant planet frequencies around AF stars.}
   {{\it Giant planets}: We report the new detection of a \msini~$= 4.51$ \Mjup~planetary companion with a $\sim$826-day period to the F6V dwarf \object{HD\,111998}. We also present new data on the 2-planet system around the F6IV-V dwarf \object{HD\,60532}.\\
{\it Spectroscopic binaries}: We also report the detections of 14 binaries with long-term RV trends and/or high-amplitude RV variations combined to a flat RV-bisector span diagram. We constrain the minimal masses and sma of these companions and check that these constrains are compatible with the stellar companions previously detected by direct imaging or astrometry for six of these targets.\\
{\it Detection limits}: We get detection limits deep into the planetary domain with 70\%~of our targets showing detection limits between 0.1 and 10 \Mjup~at all orbital periods in the 1 to $10^{3}$-day range.\\
{\it Statistics}: We derive brown dwarf (13 $\leq$ \msini~$\leq$ 80 \Mjup) occurrence rates in the 1 to $10^{3}$-day period range of $2_{-2}^{+5}$\%~and $2.6_{-2.6}^{+6.7}$\%~for stars with \Mstar~in the ranges 1.1-1.5 and 1.5-3 \Msun, respectively. As for Jupiter-mass companions (1 $\leq$ \msini~$\leq$ 13 \Mjup), we get occurrence rates in the 1 to $10^{3}$-day period range of $4_{-0.9}^{+5.9}$\%~and $6.3_{-6.3}^{+15.9}$\%~respectively for the same \Mstar~ranges. When considering the same Jupiter-mass companions but periods in the 1 to 100-day range only, we get occurrence rates of $2_{-2}^{+5.2}$\%~and $3.9_{-3.9}^{+9.9}$\%. Given the present error bars, these results do not show a significant difference with companion frequencies derived in the same domains for solar-like Main-Sequence stars.
}
   {} 

   \maketitle

\section{Introduction}\label{intro}

Since the discovery of a giant planet (hereafter GP) around a solar-type Main-Sequence (hereafter MS) star, 51 Peg \citep[][]{mayor95}, made with the \elodie~spectrograph, more than 3000 exoplanets\footnotemark~have been found, mainly using the radial velocities (hereafter RV) and transit techniques. While the first detected planets were close-in GP with a few Jupiter masses and brown dwarfs (hereafter BD), the RV method now allows the detection of Neptune-mass and mini-Neptune planets. The close-in GP are believed to have formed through the core-accretion (hereafter CA) scenario \citep{pollack96,kennedy08}, in which quickly formed massive (10-15 \ME) rocky cores accrete massive gaseous envelopes. As they represent the bulk of the planetary system mass, GPs play a key role in the shaping and final architecture of the planetary systems. They have already revealed a great diversity in terms of eccentricities, inclinations, orbital motions and especially of separations \citep{mordasini10}. The so-called ``hot planets'' (Hot Jupiters and Hot Neptunes) found at very short separations (a fraction of au) have highlighted the importance of dynamical processes such as inward migration (within a disk, or due to interactions with a third body) in the formation and dynamical evolution of planetary systems.\\

\footnotetext{Since the recent Kepler candidate statistical validation from \cite{morton16}.}

A decisive challenge is now to investigate possible correlations between these close-in ($<$5-10 au) GP and the stellar properties of their hosts, so as to better understand the impact of the stars themselves on planetary formation. For instance,the so-called ``planet-metallicity'' relation (\ie~a positive correlation between the GP occurrence rate and the stellar metallicity) is now well-established \citep[see \eg][]{fischer05,reffert14}, showing that GP preferentially form in metal-rich proto-planetary disks. 

Another important question is the impact of the central star mass on the planetary formation and evolution processes. The impact of the stellar mass \Mstar~on the core-accretion formation process and on the GP final properties is still to be fully investigated and understood. The current expectation from CA theory is that the GP frequency increases with an increasing \Mstar. \cite{kennedy08} predict a linear increase of the GP frequency for stellar masses between 0.4 and 3 \Msun, with a 6\%~frequency for \Mstar = 1 \Msun.
These predictions have been more or less validated for solar-like FGK (from F7-8V to mid-K spectral type) dwarf stars \citep[see \eg~][with a $\sim 9-10 \pm 1.5$\%~frequency of 0.3-10 \Mjup~GP at separations up to 4-5 au]{cumming08}. Concerning lower-mass M dwarfs, \cite{bonfils13} derived a smaller GP frequency of $\sim$1\%, corresponding to the expectation from CA theory \citep{laughlin04,kennedy08}. However, the increase of the GP frequency in the upper part (from $\sim$1-1.3 to $\geq$3 \Msun) of the \Mstar~range explored by \cite{kennedy08} is still to be validated for massive, AF-type Main-Sequence stars. These stars have indeed not been monitored in past RV large surveys so far, due to the specific problems they raise for precise RV computation. Massive (1.2 $\lesssim$ \Mstar~$\lesssim$ 3.5 \Msun) AF (from B9V-A0V to F6V-F7V) show far fewer spectral absorption lines and rotate faster than FGK dwarfs. These characteristics prevent one from measuring the star's RV when using the classical RV computation technique based on the cross-correlation of the stellar spectrum with a binary mask.\\

Instead, several RV surveys have focused on evolved, GK-type subgiant and giant stars off the Main-Sequence, based on the assumptions that these stars are both massive (1 $\leq$ \Mstar~$\leq$ 5 \Msun) and the descendants of AF-type MS dwarfs \citep{johnson10,bowler10}. These evolved stars also show more numerous absorption lines and slower rotation rates than AF dwarfs, allowing the classical RV computation \citep{johnson08}:
\begin{itemize}
\item Based on a $\sim$160-target sample of GK subgiants with estimated masses in the 1.2 to 2.2-\Msun~range, \cite{johnson08,johnson10} derived a GP ($\geq$ 0.8 \Mjup) frequency of $11 \pm 2$\% for separations up to 2.5 au;
\item As for red giants with estimated masses in the 1 to 5-\Msun~range, \cite{reffert14} reported an increase of the GP frequency with \Mstar~up to \Mstar~= 2.5 \Msun~(with a maximum GP rate of $\sim$15\%~in the range 1.8-2.2 \Msun), and a decrease of the GP frequency for higher stellar masses (from 2.5 up to 5 \Msun), based on a 373 GK giants sample observed for 12 years.
\end{itemize}
Such results should then confirm the GP frequency correlation to stellar mass predicted by CA theory \citep{kennedy08}, provided that the \Mstar~estimations are safe. Furthermore, a remarkable paucity of GP at short separations ($\leq$1 au) compared to solar-like dwarfs was reported for subgiant and giant stars \citep{bowler10,johnson10b,reffert14}. This trend was supposed to originate in different GP evolution mechanisms around FGK and more massive AF dwarfs, respectively \citep{bowler10}.

However, a controversy has raised in the past years on whether these evolved subgiants and giants are as massive as previously supposed and whether they are really the descendants of massive MS dwarfs. \cite{lloyd11}, based on stellar evolutionary models, and \cite{schlaufman13}, based on a galactic motion analysis, argued that evolved GK stars did not differ significantly in terms of stellar masses from FGK solar-like dwarfs. These authors argued furthermore that solar-like dwarfs are actually the predecessors of GK subgiants and giants. They finally concluded that the Hot Jupiter paucity around evolved stars was caused by tidal destruction after the star leaves the Main-Sequence, instead of early evolutionnary processes \citep{lloyd11,schlaufman13,villaver14}. This has led to an ongoing debate on stellar evolution models with \cite{johnson13a}, who argues that the differences picked up by \cite{lloyd11} between the classically adopted stellar mass distribution of evolved GP hosts and their mass distribution expected from their position in an H-R diagram have a negligible impact on their stellar mass estimates. \cite{johnson13a} maintain that these evolved stars are actually more massive than FGK dwarfs \citep{lloyd11,johnson13a,lloyd13,johnson13b,johnson14}. Besides, transit surveys have revealed a dozen or so Hot Jupiters around AF MS stars: see \eg~OGLE2-TR-L9 \citep[F3V,][]{snellen09}, \object{HD\,15082} \citep[A5V,][]{collier10} or Corot-11 \citep[F6V,][]{gandolfi10}. In this context, AF MS stars can become targets of choice to investigate the impact of stellar mass on both the GP frequency and GP period distribution.\\

We developed ten years ago the Software for the Analysis of the Fourier Interspectrum Radial-velocities \citep[\safir,][]{galland05a}, dedicated to the RV computation of AF, MS stars. For each target, instead of correlating the stellar spectrum with a classical binary mask to measure the RV, \safir~correlates the spectrum with a reference spectrum built from the median of all the spectra acquired on this target and properly shifted to a common wavelength scale. This procedure has rapidly proved its ability to detect BD as well as GP around AF, MS stars \citep[see \eg][]{galland05b,galland05a,galland06}. Since 2005, we have initiated and carried out two feasibility surveys dedicated to the search for close-in (up to a few au from the host star) GP and BD around AF, MS stars: one in the northern hemisphere with the \sophie~fiber-fed spectrograph \citep[][]{bouchy06} on the 193 cm telescope at the Observatoire de Haute-Provence (OHP, France), and the other one in the southern hemisphere with the High-Accuracy Radial Velocity Planet Searcher \citep[\harps;][]{pepe02} on the 3.6m ESO telescope at La Silla Observatory (Chile). We focus here on the \harps~southern stars. \cite{lagrange09} reported on the results of the three first years of this \harps~survey (2005-2008). These authors showed that most of the 185 targets were intrinsically RV variable, and characterized the RV variability. They also made a first assessment of the survey sensitivity to substellar companions, achieving detection limits at short periods ($\le$ 100 days) deep into the planetary domain for most of the targets, despite early spectral type and/or high RV jitter. Finally, the discovery of a 2-GP system in a mean-motion resonance around the F6IV-V dwarf \object{HD\,60532} was reported in \cite{desort08}.

\begin{figure*}[ht!]
  \centering
\includegraphics[width=0.9\hsize]{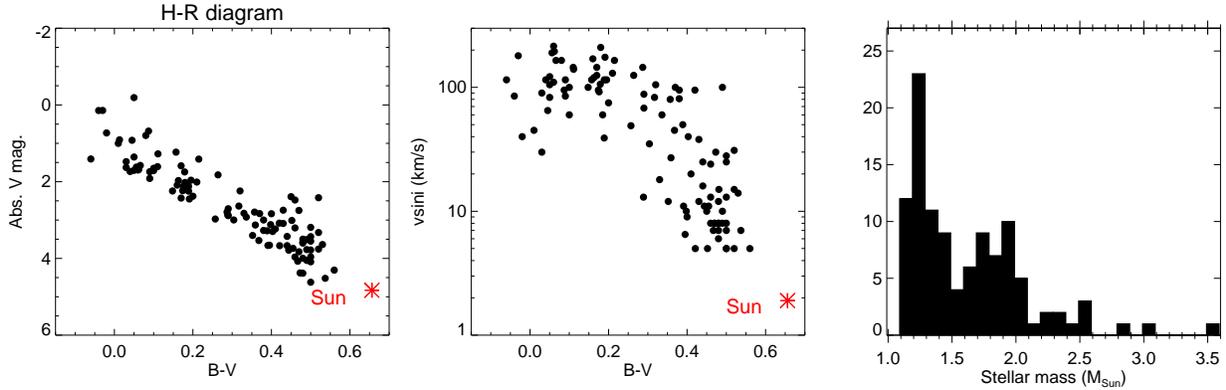}
\caption{Our A-F sample properties. {\it Left}: HR diagram of our sample, in absolute $V$-magnitude vs \bv. The Sun is displayed ({\it red star}) for comparison. {\it Middle}: \vsini~vs \bv~distribution. {\it Right}: Mass histogram.} 
       \label{sample}
\end{figure*}

Following this feasibility survey, we carried on a \harps~survey from 2008 to 2011 with a restricted sample of 108 of the already observed targets, and extended further the observations in 2013-2014 for the most interesting targets. We followed a similar procedure for our \sophie~observations from 2006 to 2013, and introduced two new GP (one candidate GP and another confirmed one) detected in the course of this \sophie~survey in \cite{desort09} and \cite{borgniet14}, respectively. The present paper is dedicated to the description and analysis of the \harps~survey, and a forthcoming one will present the \sophie~survey. We detail our sample properties, our observations, our observables and the criteria we use to characterize RV variability in Sect.~\ref{data}. We introduce the newly reported and/or further characterized GP in Sect.~\ref{planets}, as well as the detected stellar binaries in Sect.~\ref{binaires}. We characterize the RV intrinsic variability of our targets in Sect.~\ref{intr_var}. Finally, we introduce in Sect.~\ref{limdets} the detection limits we achieved on our targets, and we derive a first estimation of the close-in GP and BD occurrence rates for AF MS stars, before concluding in Sect.\ref{conclu}.

\section{Description of the survey}\label{data}

\subsection{Sample}

Our \harps~sample is made up of 108 MS stars with spectral types (ST) in the B9V to F9V range. The B9V-A0V cutoff roughly corresponds to the earlier ST for which our detection limits are expected to fall into the planetary domain given our targets \vsini~and the results of our feasibility survey \citep{lagrange09}. The F6V-F9V cutoff more or less corresponds to the earliest ST for which the classical masking technique can be used to compute the RV.

Our sample is limited to nearby stars, with a distance to the Sun less than 67 pc for B9V-A9V dwarfs and less than 33 pc for F0V-F9V dwarfs. The difference in distance between A and F-type stars is meant to keep roughly the same number of stars of each ST range in our sample.

We also remind that we first removed from the feasibility survey sample any previously known spectroscopic binary, as well as close visual binaries with a separation under 5 arcseconds. Finally, we removed known variable stars of $\delta$ Scuti or $\gamma$ Doradus type, as such pulsators induce high-amplitude RV variations over periods ranging from a few hours to a few days that undermine any search for companions (unless for targets of peculiar interest such as \object{HR\,8799}). We also removed known Ap-Am stars that show spectral anomalies and that are often associated to binary systems. This removes a significant number of late A to early F-type stars at the crossing of the instability strip on the Hertzsprung-Russell (HR) diagram (Fig.~\ref{sample}).

We ended up with 108 stars with \bv~in the range $-$0.04 to 0.58. All of them rotate faster than the Sun (Fig.~\ref{sample}). The mass distribution extends from 1 to 3.6 \Msun~but mostly covers the 1.1-2.7 \Msun~range (Fig.~\ref{sample}). The list of our targets, together with their main relevant properties, is provided in Appendix~\ref{whole_survey}.

\begin{figure*}[ht!]
  \centering
\includegraphics[width=0.9\hsize]{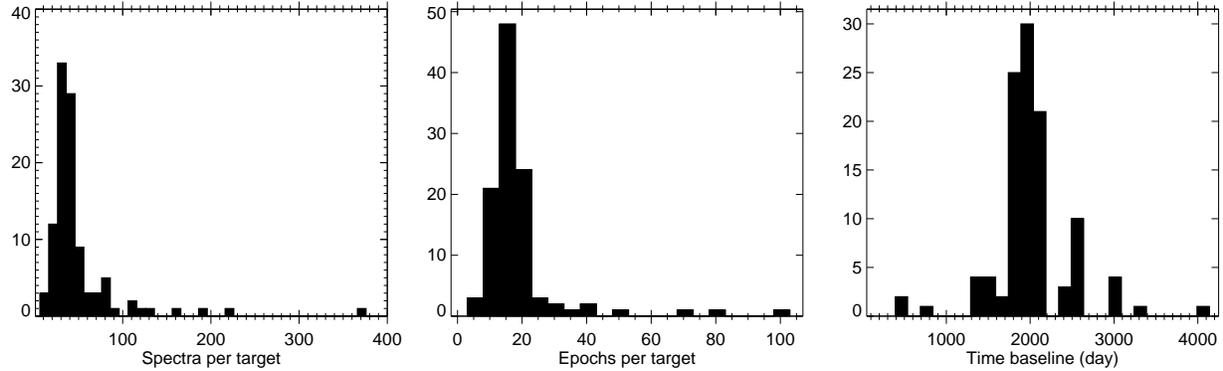}
\caption{Observation summary. {\it Left}: Histogram of the spectrum number per target. {\it Middle}: Histogram of the separate observation epoch number per target. {\it Right}: Histogram of the time baselines.} 
       \label{sample_obs}
\end{figure*}

 \subsection{Observations}

We observed our 108 targets with \harps~mainly between August 2008 and August 2011, in continuation of the survey described in \cite{lagrange09}, that already covered 2.5 years (August 2005 to January 2008). As most of our targets were already part of this previous survey, this allowed to double their observation time baseline. We finally acquired a few additional spectra lately (in 2014) on our most interesting targets. We acquired the \harps~spectra in the 3800-6900~\AA~wavelength range, in high-resolution mode ({\it R} $\simeq 115000$). We adapted the exposure times (between 60s and 900s, depending on the stars magnitudes and on the observing atmospheric conditions) in order to reach a high signal-to-noise ratio (hereafter S/N) of 260 on average at 550 nm. We made most of the exposures in the fibre spectroscopy mode, for which the \harps~A fiber is centered on the target and a dark on the B fiber. This mode is convenient for targets much brighter than the background, such as our AF targets. As for the few spectra acquired in 2014, we acquired them in the simultaneous Thorium mode. In that case, the first \harps~fiber is centered on the target, while the second is fed by a Thorium lamp. The Thorium spectrum acquired simultaneously to the stellar one allows one to follow and correct for the drift of the instrument induced by local temperature and pressure variations.\\

The observing strategy is the same as described in \cite{lagrange09}. For each pointed star, we recorded at least two consecutive spectra (each pointing is hereafter referred as an ``epoch''). For each target, we also tried to record data on several consecutive nights to estimate the short-term jitter and detect potential high-frequency variations induced by pulsations. Our typical observation time baseline is $\sim$1900 days ($\sim$5.2 years) per target (Fig.~\ref{sample_obs}). The number of spectra acquired per target depends on its interest and on the potential hint of a companion. For stars with no companion detection, the median number of acquired spectra $N_{\rm m}$ is in the 20 to 50-spectra range, roughly corresponding to 15 to 25 different epochs (Fig.~\ref{sample_obs}). The RV computation is described Sect.~\ref{observables}. We made a first selection of the spectra based on two criteria:
\begin{enumerate}
\item the S/N at $\lambda = 550$ nm must be greater than 80 to eliminate spectra acquired with poor observing conditions and lower than 380 to avoid saturation;
\item the atmospheric absorption must be kept to a minimum. We defined an absorption parameter, described in \cite{borgniet14}, that allows us to estimate the absorption by the atmosphere during the observations and thus the quality of the observations.
\end{enumerate}

\subsection{Observables}\label{observables}

 \subsubsection{Radial velocities (RV)}
 
We computed the RV with our dedicated \safir~software. \safir~and the method used to compute RV for early-type, fast-rotating stars are described in \cite{chelli00} and \cite{galland05a}. We remind briefly its principle here. We use the 2D ``ed2s'' spectra firstly reduced by the \harps~Data Reduction System (DRS) pipeline as the input data to our software. For each star of our sample, we build a first estimate of our reference spectrum by computing the median of the acquired spectra. At this step, we also compute the $\chi^2$ of each spectrum compared to the reference spectrum so as to assess its quality. When the $\chi^2$ is found to be over 10, the spectrum is not kept in the estimation of the reference. Such a case happens rarely and can be originating in bad observing conditions (although this is already taken into account during the selection of the spectra based on the absorption, see above), in technical problems or in line deformations induced by a double-lined spectroscopic binary (hereafter SB2). 
We then compute the correlation (in the Fourier domain) between the reference spectrum and each spectrum to determine a first estimate of the RV. We rebuild a final reference spectrum by computing the median of all the spectra once shifted from the first RV measurement. The final RV value is obtained by correlating again each spectrum with the final reference spectrum.

\subsubsection{Bisector velocity span (BIS) and full width at half maximum (FWHM) of the spectra cross-correlation function (CCF)}
 When possible, we computed for each target the cross-correlation function (hereafter CCF) of the spectrum and the corresponding bisector and bisector velocity span \citep[hereafter BIS; see \eg][]{queloz01,galland05b,galland06}. The CCF computation with \safir~is possible for stars with ST later than A0 and with \vsini~typically $\leq 150$\kms, \ie~for stars that have a sufficient number of spectral lines and that are not too rotationally-broadened. The \safir~CCF are obtained by cross-correlating each spectrum with an automatically built binary mask based on the deepest and not blended lines of the reference spectrum. The uncertainty associated to the BIS depends on the \vsini~and on the number of spectral lines available for the CCF computation \citep{lagrange09}.

The BIS and the full width at half maximum (FWHM) of the CCF are very good diagnosis of stellar activity due to magnetically active structures (dark spots and bright faculae) or high-frequency pulsations. Yet, for a low \vsini~(of the order or smaller than the instrumental spectral resolution), the flux variation induced by a magnetically active structure and the corresponding global CCF distortion will not have a significant effect on the BIS compared to the RV \citep[see \eg][]{desort07}. Indeed, if the \vsini~is smaller than the instrumental resolution, the spectral lines are not resolved and the BIS variations are negligible compared to the RV ones. On the contrary, the activity-induced effect on the FWHM should always be noticeable \citep[of the same order of or larger than the RV variations, see][]{dumusque14}. For a higher \vsini~($\geq$ 5-6 \kms), the activity-induced BIS variations will always be noticeable and will be larger than the FWHM variations \citep{desort07,dumusque14}. Here, as most of our targets have a high \vsini, the BIS will be our main proxy for stellar activity (see below). 

\subsubsection{Chromospheric emission}

\safir~also allows to measure the chromospheric emission in the Calcium (Ca) H and K lines, expressed either in the $S$-index which gives the ratio of the flux measured in the core of the Ca H and K lines by the flux measured in two continuum bands on either side (blue and red) of the Ca lines; or the \rhk, which gives the log of the $S$-index to which the photometric emission in the Ca lines has been substracted (therefore keeping only the chromospheric emission). The \rhk~increases with the active region surface coverage or filling factor\footnotemark~and is therefore commonly used as a stellar magnetic activity proxy for solar-like stars \citep[see \eg][]{dumusque12,meunier13,santos14}. When available, we use here the \rhk~of our targets as an activity proxy complementary to the BIS.

\footnotetext{In the case of the Sun, for which the bright faculae are much larger than the dark spots, the \rhk~increases linearly with the bright facula coverage and quadratically with the dark spot coverage, according to \cite{shapiro14}.}

\subsection{Classification of the RV variable targets}\label{classif}

As in \cite{lagrange09}, we consider a target as a RV variable if its total RV amplitude is larger than six times the mean RV uncertainty and if its RV standard deviation (equivalent to the RV rms) is larger than twice the RV mean uncertainty. A fully detailed description of how the combined use of RV, CCF and BIS allows to distinguish between RV variations induced either by companions or by intrinsic stellar activity is provided in \cite{lagrange09} for AF MS stars. We remind here only the main points of this classification:
\begin{enumerate}
\item In the case of RV variations induced by stellar magnetic activity (\ie~dark spots and bright faculae), the RV and BIS are anti-correlated if the \vsini~is larger than the instrumental spectral resolution. In a (RV,~BIS) diagram, the BIS values are arranged either in a linearly decreasing function of the RV, or in an inclined ``8''-shape, depending on the \vsini, the stellar inclination and on the active structure configuration at the stellar surface \citep[see \eg][]{desort07,boisse11,lagrange13}. More complex patterns can be found in the case of multiple large structures.
\item In the case of stellar pulsations, the BIS shows important variations over a much larger range than the RV. The BIS and RV variations are not correlated any longer. In the (RV,~BIS) diagram, the BIS values show a vertical spread.
\item In the case of BD or GP companionship as the source of RV variations, the CCF is not distorted as in the case of active magnetic structures and is only shifted in RV. In the case of double-line spectroscopic binaries (hereafter SB2), the CCF is strongly distorted due to the two spectra overlapping, inducing strong RV as well as BIS and FWHM variations. However, we remind that we removed any known SB2 from our sample, and that the SB2 we detected in \cite{lagrange09} were also removed. Hence, the present sample should not contain any SB2 with periods up to several hundreds days. Only SB2 with periods of a few to a few tens years should still be present in our sample. Given our typical time baselines, such ``long-period'' SB2 binaries should be unresolved spectroscopically, \ie~the RV shift between the two components stays smaller than the global FWHM over the observation time baseline. In this case, the CCF distortions might be difficult to detect and only FWHM variations correlated to the RV variations could be seen. Moreover, the BIS would not show significant variations (hence an apparently flat (RV,~BIS) diagram), and the RV-FWHM correlation would be the only criterion to distinguish this case from a single-lined spectroscopic binary or SB1 \citep{santerne15}. A SB1 does not induce CCF distortions nor FWHM or BIS variations, whereas the RV are strongly variable. The corresponding (RV,~BIS) diagram is then ``flat''. We use the same diagnosis in the case of companions of lower mass (BD and GP), though in this case the RV variations are of lower amplitude. 
\item Finally, if the star is both member of a binary system and active, the (RV,~BIS) diagram will be composite, with both an horizontal spread induced by the companion and an inclined or vertical spread induced by active structures or pulsations, respectively. The origin of the dominant spread depends on the activity strength and on the companion properties.
\end{enumerate}
We display the main results in terms of observations, observables and RV analysis for all our targets in Appendix~\ref{whole_survey}.

\begin{figure*}[t!]
  \centering
\includegraphics[width=0.85\hsize]{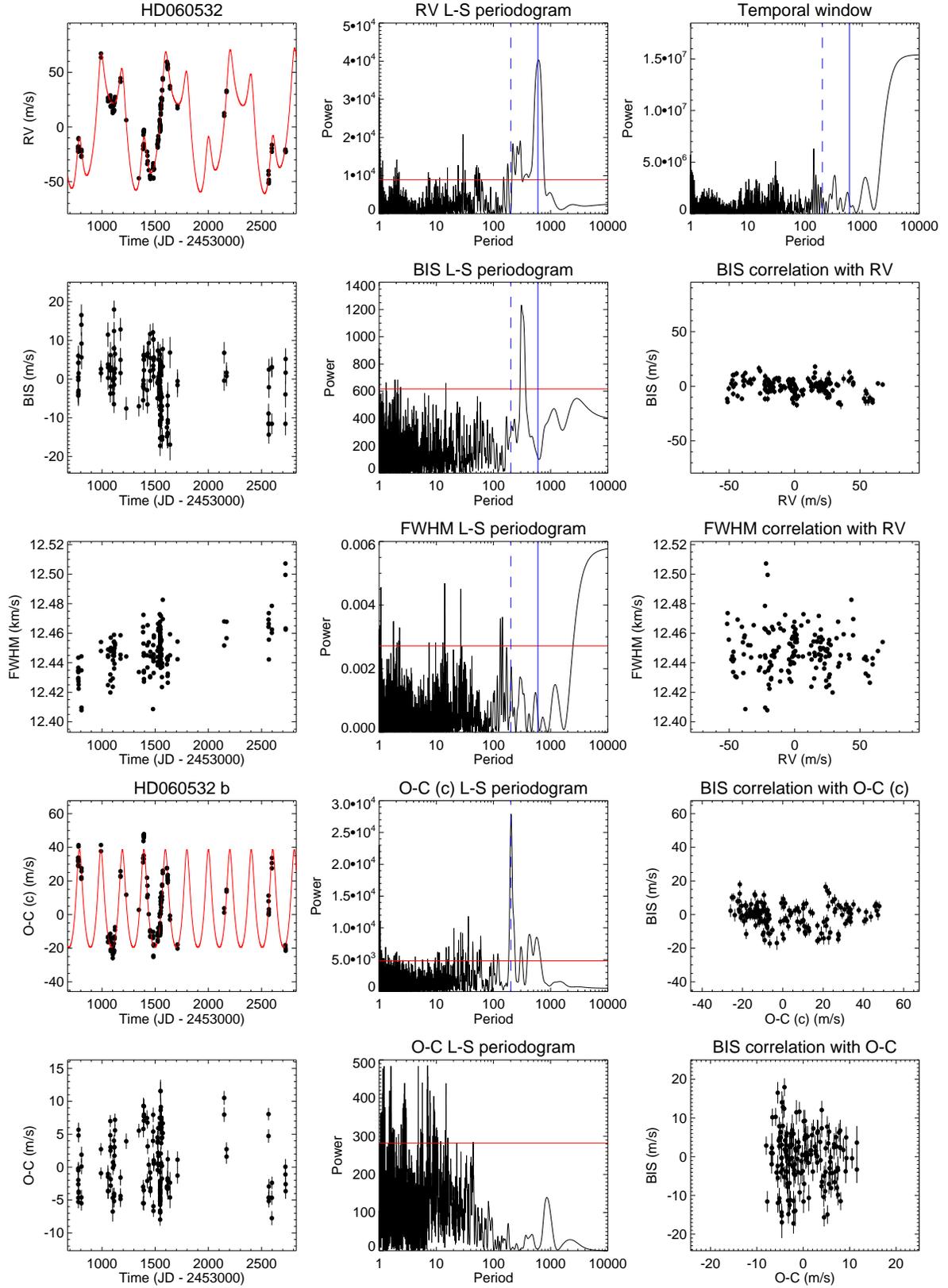}
\caption{\object{HD\,60532} spectroscopic data. {\it Top row}: \object{HD\,60532} RV time series ({\it left}), Lomb-Scargle periodogram of the RV ({\it middle}) and temporal window of the observations ({\it right}). The Keplerian fit is superimposed ({\it red solid line}) to the RV. On the periodogram, the false-alarm probability (FAP) at $1\%$ ({\it solid line}) is indicated in red; the planet periods are indicated in blue. {\it Second and third rows}: BIS and FWHM time series, corresponding Lomb-Scargle periodograms and correlations with the RV data. {\it Fourth row}: RV residuals from planet c, corresponding Lomb-Scargle periodogram and correlation with BIS. The planet b Keplerian fit is superimposed to the residuals, and planet b period is superimposed in blue on the periodogram. {\it Fifth row}: RV residuals from the 2-planet Keplerian fit, corresponding Lomb-Scargle periodogram and correlation with BIS.}
       \label{hd60532}
\end{figure*}

\section{Giant planet detections}\label{planets}
The present data set reveals three GP companions to two of our targets. Two GP were already known to orbit around \object{HD\,60532} \citep{desort08}. We add new RV data to this system. Then, we report the detection of a new GP orbiting around the F6V dwarf \object{HD\,111998}. We detail the stellar characteristics of these two targets in Table~\ref{stellparam}.

\subsection{The 2-GP system around \object{HD\,60532}}

\subsubsection{System characteristics} 
This system was discovered and described in \cite{desort08}. \object{HD\,60532} is a F6IV-V star hosting a system of two GP with minimal masses of $1.03 \pm 0.05$ and $2.46 \pm 0.09$ \Mjup~and with semi-major axis (sma) of $0.759 \pm 0.01$ and $1.58 \pm 0.02$ au, respectively. The \object{HD\,60532} system shows a 3:1 mean motion resonance (hereafter MMR) stable at the Gyr timescale \citep{desort08,laskar09}. We continued to follow \object{HD\,60532} during our survey, adding 28 spectra to the 147 already acquired at the beginning of 2008, and expanding consequently the observation time baseline from 2 to 5.5 years (1949 days). We performed a new fit of the RV data set (\ie~175 spectra) with a Keplerian 2-planet model to test the impact of the new data on the orbital parameters of the system. For that purpose, we used the {\it yorbit} software \citep{segransan11}. This software uses a Levenberg-Marquardt algorithm to fit the RV data with Keplerian models, after selecting the values with a genetic algorithm. We found a solution very close to the best model given in \cite{desort08}, with non-significant differences. This strongly strengthens the detection of the two planets. We give the orbital parameters of the system obtained with the new Keplerian fit in Table~\ref{planetparam}. The RV data along with the 2-planet Keplerian fit are displayed in Fig.~\ref{hd60532}. 

The RV Lomb-Scargle periodogram is strongly dominated by the signal of planet c and its aliasing. We computed the Lomb-Scargle periodogram using the fast-algorithm method of \cite{press89}, that gives the power spectrum of the RV data \citep[according to the definition of][]{scargle82} versus the period range. We finally display in Fig.~\ref{hd60532} the RV residuals from planet c only, and from the 2-planet Keplerian fit. We note that the planet b signal dominates the periodogram of the RV residuals from planet c, and that the remaining short-period signals in the final RV residual periodogram is most likely induced by low-intensity stellar activity, given the shape of the (RV residuals,~BIS) diagram (Fig.~\ref{hd60532}).

\subsubsection{Remarks on the line profiles} 
We display all the \object{HD\,60532} relevant spectroscopic data (RV, BIS, FWHM) in Fig.~\ref{hd60532}. As stated in \cite{desort08}, the (RV,~BIS) diagram is flat (with an RV amplitude of $\sim$120 \ms~and a BIS amplitude of $\sim$35 \ms), with no correlation. However, the BIS Lomb-Scargle periodogram shows a strong peak at 309 days. Such a periodicity is hardly noticeable when looking at the BIS time series (Fig.~\ref{hd60532}). Noting that a similar peak at $\sim$320 days is also present in the the observation temporal window, we conclude that the $\sim$300-day BIS peak most probably originates in temporal sampling effects. As for the periodogram of the FWHM, the main peaks located at $\sim$30 and $\sim$140 days are also present in the temporal window, also hinting towards sampling effets. The FWHM and, less significantly, the BIS show a long-term trend of low amplitude over the observation time baseline, that may be indicative of long-term stellar activity.

\subsubsection{Additional remarks}
The third version of the Geneva-Copenhagen Survey \citep[hereafter GCS III,][]{holmberg09} gives an age of 2.8 Gyr for \object{HD\,60532}, based on the Padova stellar evolution model \citep[][]{holmberg07}. A new analysis of the GCS led to a similar age estimation of 2.4 Gyr \citep[][]{casagrande11}. However, the GCS often finds ages much older than they actually are in the case of young stars. Yet \object{HD\,60532} is not known to be a member of a young association. Its ({\it UVW}) space velocities ($-37$, $-49$, $-3$ \kms) are not compatible with the ones of the known young moving groups listed in \cite{torres08} or \cite{nakajima12}. Furthermore, its luminosity class (IV-V) and relatively high radius estimation \citep[$R_{\star}$ = 2.57 \Rsun,][]{allende99} do not point towards a young star.

Orbital analysis and simulations of the 3:1 MMR in the \object{HD\,60532} planetary system favour a small inclination of the system \citep[$i \sim$20\degr,][]{laskar09,sandor10}. The true masses of \object{HD\,60532} planets would then be increased by a factor $\sim$3 compared to their minimal masses, \ie~$\sim$3.2 and $\sim$7.5 \Mjup~for planets b and c, respectively.

 \cite{mcdonald12} reported a weak IR excess around \object{HD\,60532} of $\sim$30\%~in flux in average between 3.5 and 25 $\mu$m through SED fitting. Such a weak excess may be induced by circumstellar dust; however \cite{mcdonald12} does not propose a size for the excess emission. No circumstellar dust has been resolved so far around \object{HD\,60532}.

\subsection{A new GP around \object{HD\,111998}}

\subsubsection{RV data} 
\begin{figure*}[t!]
  \centering
\includegraphics[width=0.85\hsize]{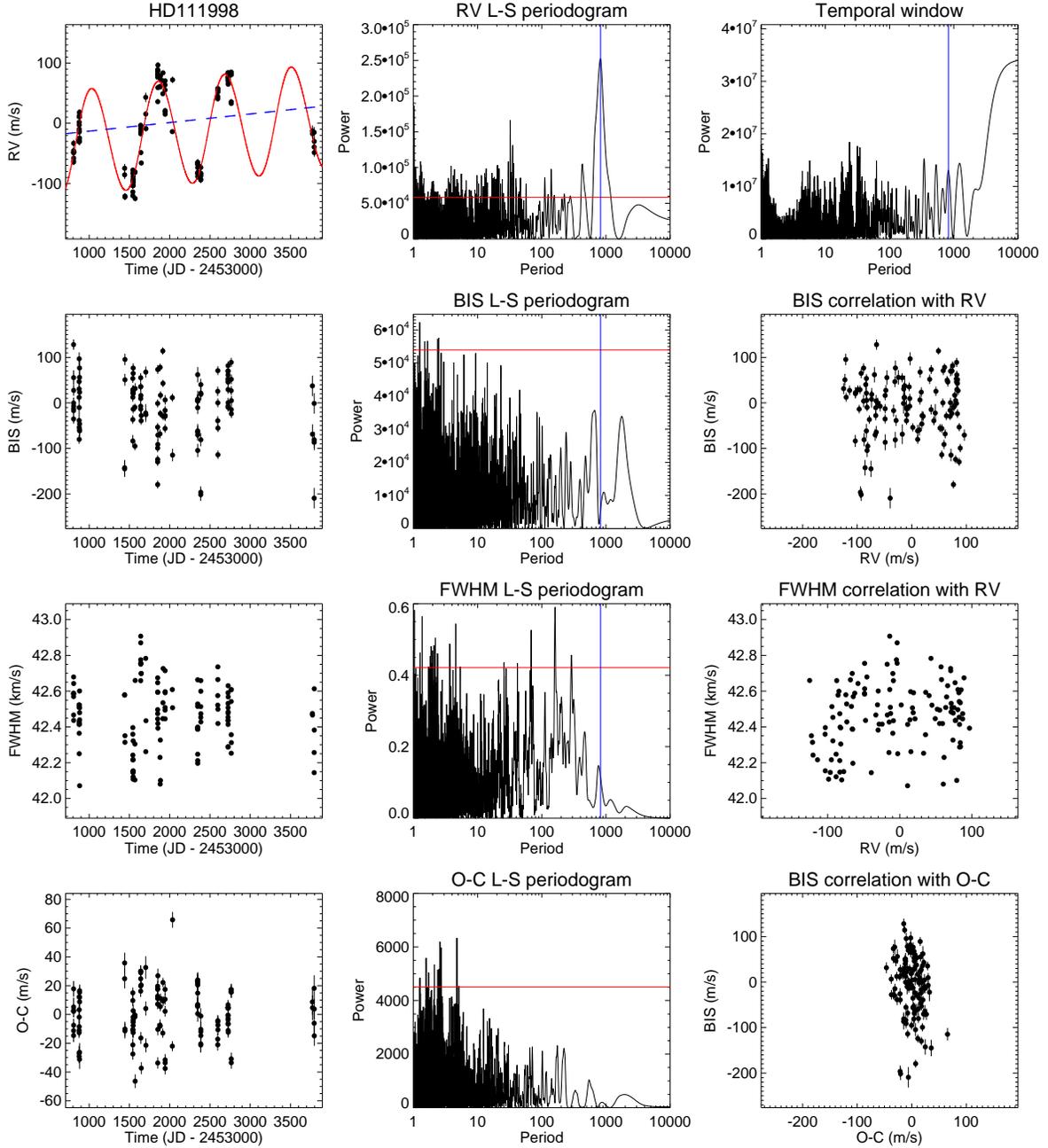}
\caption{\object{HD\,111998} spectroscopic data. {\it Top row}: \object{HD\,111998} RV time series ({\it left}), RV Lomb-Scargle periodograms ({\it middle}) and temporal window of the observations ({\it right}). The Keplerian fit is superimposed ({\it red solid line}) to the RV, as well as the linear trend ({\it dashed blue line}). On the RV periodogram, the FAP at $1\%$ ({\it solid line}) is indicated in red; the planet period is indicated in blue. {\it Second and third rows}: BIS and FWHM time series, corresponding Lomb-Scargle periodograms and correlations with the RV data. {\it Fourth row}: Residuals of the fit as a function of time, periodogram of the residuals and BIS correlation with the residuals.}
       \label{hd111998}
\end{figure*}

We obtained 127 high S/N \harps~spectra (the average S/N is 275) on 38 Vir (\object{HD\,111998}, \object{HIP\,62875}, F6V), covering a 2989-day (8.2 year) time baseline. The RV data show a clear periodic signal with a peak-to-peak amplitude of $\sim$222~\ms, and a dispersion of 66.4~\ms. These values are well above the 4.7~\ms~average uncertainty on the RV (accounting for photon noise and instrumental stability). The RV data and their Lomb-Scargle periodogram are given in Fig.~\ref{hd111998}. The RV periodogram shows several peaks above the 1\%~FAP, with the highest by far at a period of about 820 days (which we will attribute to a planet, see below). Another peak at 422 days appears to be an alias of the $\sim$820-day period. Finally, we attribute the 32-day peak to temporal sampling effects, noting that a similar peak is present in the temporal window (Fig.~\ref{hd111998}). 

\subsubsection{Line profile data} 
We show in Fig.~\ref{hd111998} the BIS and FWHM as a function of time, as well as their Lomb-Scargle periodograms, and the (RV,~BIS) and (RV,~FWHM) diagrams. Although the BIS shows high-amplitude variability (with a dispersion of $\sim$67~\ms~and a peak-to-peak amplitude of $\sim$337.5~\ms), it does not show any significant temporal periodicity in the 100 to 2000-day range. The BIS periodogram shows power at high frequencies (at periodicities in the 1 to 7-day range). Moreover, there is no correlation between the RV and the BIS (Pearson's correlation coefficient $\sim$0.03) despite the large \vsini~(28 \kms). The (RV,~BIS) diagram clearly shows first a main horizontal (``flat''), long-term spread (pointing towards the presence of a companion), and then a secondary, short-term, slope that corresponds to low-intensity activity. This is better seen when looking at snapshots of the RV and BIS data (Fig.~\ref{snap111998}). The FWHM of the CCF shows variability both in the 1 to 7-day range and in the 100 to 400-day range. However, the FWHM periodogram does not exhibit any significant power in the 400 to 2000-day range, and there is no significant correlation between the RV and FWHM data (Pearson coefficient of $\sim$0.2). Finally, \object{HD\,111998} does not show any significant emission in the Calcium H and K lines. We find a mean \rhk~of -4.69, in agreement with previous measurements for \object{HD\,111998} ($<$~\rhk~$>$ = -4.77, -4.44 according to \cite{pace13} and \cite{murgas13}, respectively).

\subsubsection{Origin of the RV variations} 
We investigate here the possible origins of the $\sim$820-day periodic RV variation. Stellar pulsations are highly improbable as a periodicity of more than 800 days is far larger than those of any type of known pulsations for Main-Sequence stars. They would also induce a much larger variability in the BIS at much longer periods than those observed. We also exclude activity-induced variability as dark spots or bright faculae would induce signals with periods of a very few days. Indeed, given the star \vsini~($\sim$28~\kms) and the radius estimations from \cite{allende99} and \cite{pasinetti01} (1.45 and 1.0 \Rsun, respectively), the stellar rotational period would be under three days if the star is seen edge-on, and would be even shorter if the star is seen inclined. This is clearly not compatible with the observed $\sim$820-day period. Finally, we also rule out longer-term effects of stellar magnetic activity (\ie~stellar cycles): with this \vsini~level, such effects on the RV would induce correlated high-amplitude variations on the BIS and FWHM at similar periods, which is not the case. We therefore attribute the $\sim$820-day RV periodic variation to the presence of a companion.

\subsubsection{Keplerian fit} 
We fitted \object{HD\,111998} RV data with a 1-planet Keplerian model and a linear trend (with a $5.2 \pm 1.3$ \msy~slope) simultaneously, using the {\it yorbit} software. Adding this slight linear trend to the Keplerian model allows to significantly reduce the rms of the residuals (by 1.1 \ms, see below). The best solution is a companion on an almost circular orbit ($e = 0.03 \pm 0.04$), with a period of $825.9 \pm 6.2$ days and a semi-amplitude $K = 87.6 \pm 3.4$ \ms. Taking the stellar mass and its error bar into account (\Mstar~$ = 1.18 \pm 0.12$ \Msun), this corresponds to a GP with \msini~$= 4.51 \pm 0.50$ \Mjup. To test the potential impact of the RV variability observed at high frequencies (for periods of one day or less), we also fitted the RV data averaged over one, two and five days with a 1-planet Keplerian model. In all cases, we obtain the same orbital parameters as with the original RV data with differences lower than the uncertainties, showing the reliability of our model. We display the best-fit (Keplerian+linear) model superimposed on the \harps~RV in Fig.~\ref{hd111998}. The orbital parameters of the \object{HD\,111998} planetary system are listed in Table~\ref{planetparam} along with their $1\sigma$~uncertainties.\\

The slight long-term RV trend that we fitted simultaneously to the Keplerian model is not seen in the FWHM data, though a similar trend may be present in the BIS data but with an opposite slope. This would imply a stellar origin. The effect of a distant binary companion to \object{HD\,111998} might be another explanation to this RV trend. 38 Vir is associated to a \rosat~X-ray source (\cite{haakonsen09}, which would be consistent with an M-type distant companion. However, this remains quite speculative at this stage, as the slope of the linear trend is very small and we do not have additional information about this potential wide binary companion. 

\begin{figure}[ht!]
  \centering
\includegraphics[width=0.95\hsize]{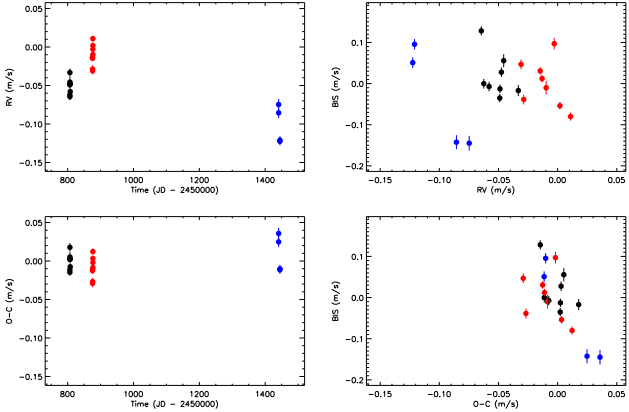}
\caption{{\it Top}: focus on \object{HD\,111998} RV data over 800 days ({\it left}) at three distinct epochs ({\it black, red and blue dots}); and corresponding (RV,~BIS) diagram ({\it right}). {\it Bottom}: the same for \object{HD\,111998} RV residuals.}
       \label{snap111998}
\end{figure}

\subsubsection{Interpretation of the residuals}
 We display the residuals of our best Keplerian fit as well as their Lomb-Scargle periodogram in Fig.~\ref{hd111998}. The residuals show a greater dispersion ($\sigma_{\rm O-C} \simeq 17.4$ \ms) than the mean RV uncertainty ($\sim$4.7 \ms). They do not show any significant power at periodicities above 20 days. The power is mainly located at high frequencies, with two groups of peaks around 2.5 and 5 days. As these peaks correspond to low-amplitude peaks that are present in the BIS or FWHM periodograms, we conclude that they are most likely induced by stellar activity. The 2.5-day peak might correspond to the stellar rotational period, if the star is seen nearly edge-on (see above). We note that the correlation coefficient between the RV residuals and the BIS is higher ($\sim$0.2) than the RV-BIS correlation ($<$0.1), even if not significant in terms of a correlation (Fig.~\ref{hd111998}). When looking at the (RV residuals,~BIS) diagram for several data snapshots, the inclined or vertical spread corresponding to low-level activity or low-amplitude pulsations is clear (Fig.~\ref{snap111998}).

\subsubsection{Additional remarks} 

\object{HD\,111998} age in the GCS III catalog \citep[][]{holmberg09} is 1.9 Gyr. However, the GCS re-analysis of \cite{casagrande11} assigns a much younger age of 600 Myr to \object{HD\,111998}. This new age estimation would be in agreement with a Hyades cluster membership, as reported by \cite{eggen82}. However, such a Hyades membership for \object{HD\,111998} has not been confirmed since.

\object{HD\,111998} is not currently known to show a clear IR excess that would point towards the presence of a debris disk. However, this target has not been included yet in a specific survey dedicated to the search for IR excesses to our knowledge. Given the detection of such a GP companion and its revised age estimation, \object{HD\,111998} should be a target of interest for future searches for debris disks, even though there is no clear correlation between cold debris disks and GP detected by RV. More systems need to be investigated to test an hypothetical correlation between RV GP and debris disks. For example, the \object{HD\,113337} system hosts both a debris disk \citep{chen14} and at least one RV GP \citep{borgniet14}.

\renewcommand{\arraystretch}{1.25}
\begin{table*}[t!]
\caption{Stellar properties of our targets with detected or possible GP.}
\label{stellparam}
\begin{center}
\begin{tabular}{l c c c}\\
\hline
\hline
Parameter    & Unit  & \object{HD\,60532}                  & \object{HD\,111998}                 \\         
\hline
Spectral type&       & F6IV-V\tablefootmark{a}               & F5V\tablefootmark{b}            \\     
$V$          &       & 4.44\tablefootmark{c}              & 6.11\tablefootmark{c}                        \\                         
$B-V$        &       & 0.52\tablefootmark{c}              & 0.49\tablefootmark{c}                  \\                     
\vsini       & [\kms]& 8.\tablefootmark{d}                & 20.\tablefootmark{d}                \\      
$\pi$        & [mas] &  $39.53 \pm 0.27$\tablefootmark{e} & $30.55 \pm 0.30$\tablefootmark{e}\\ 
$[$Fe/H$]$   &       &  -0.26\tablefootmark{f}           &   0.07\tablefootmark{f}          \\   
$T_{\rm eff}$  &   [K] &  $6245 \pm 80$\tablefootmark{g}   & $6557 \pm 96$\tablefootmark{g}  \\ 
$\log{g}$    & [dex] &$3.8 \pm 0.07$\tablefootmark{h}      & $4.19 \pm 0.11$\tablefootmark{h}  \\  
\Mstar       &[\Msun]&$1.50 \pm 0.09 $\tablefootmark{h}    & $1.18 \pm 0.12$\tablefootmark{h}  \\     
Radius       &[\Rsun]& $2.57 \pm 0.12$\tablefootmark{h}    & $1.45 \pm 0.07$\tablefootmark{h}  \\  
             &       & 1.5 - 3.2\tablefootmark{i}          &     1.00\tablefootmark{i}       \\  
Age          & [Gyr] &  $2.8_{-0.6}^{+0.1}$\tablefootmark{f} & $1.9_{-0.7}^{+0.6}$\tablefootmark{f} \\
             &       &  $2.4_{-0.2}^{+0.7}$\tablefootmark{g} & $0.6_{-0.5}^{+1.6}$\tablefootmark{g}       \\  
\hline
\end{tabular}
\tablefoot{
\tablefoottext{a}{\cite{gray06}}
\tablefoottext{b}{\cite{malaroda75}}
\tablefoottext{c}{\cite{esa97}}
\tablefoottext{d}{Estimation from the SAFIR software.}
\tablefoottext{e}{\cite{vanleeuwen07}}
\tablefoottext{f}{\cite{holmberg09}}
\tablefoottext{g}{\cite{casagrande11}}
\tablefoottext{h}{\cite{allende99}}
\tablefoottext{i}{\cite{pasinetti01}}
}
\end{center}
\end{table*}

\renewcommand{\arraystretch}{1}
\renewcommand{\arraystretch}{1.25}
\begin{table*}[t!]
\caption{Best orbital solutions.}
\label{planetparam}
\begin{center}
\begin{tabular}{l c c c}\\
\hline
\hline
          Parameter               & HD60532b         & HD60532c                & HD111998b                   \\    
\hline
          $P$  [day]              &$201.9 \pm 0.3$   &$600.1 \pm 2.4$          & $825.9 \pm 6.2$                  \\                  
          $T_0$   [BJD-2453000]   &$1594.7 \pm 2.8$  &$1973.0 \pm 100.1$       & $2490.2 \pm  177.3$              \\                      
          $e$                     &$0.26 \pm 0.02$   &$0.03 \pm 0.02$          & $0.03 \pm  0.04$                \\                          
          $\omega$  [deg]         &$-3.7 \pm 5.7$    &$179.8 \pm 58.7$         & $-87.3 \pm 77.7$                 \\                           
          $K$  [\ms]              &$29.1 \pm 0.9$    &$46.1 \pm 1.0$           & $87.6 \pm 3.4$                   \\                          
  \hline
          Add. linear trend [\msy] & 0                & -                      & $5.2 \pm 1.3$                    \\
  \hline
          $N_{\rm m}$              & 175              & -                       & 124                              \\                 
          $\sigma_{O-C}$ [\ms]     & 4.66 (27.56)~$^{\star}$  & -                 & 17.35 (67.74)~$^{\star}$            \\                      
          reduced $\chi^{2}$      & 4.44 (25.4)~$^{\star}$   & -                 & 4.18 (15.91)~$^{\star}$           \\                  
  \hline
          $m_{P}\sin{i}$ [\Mjup] &$1.06 \pm 0.08$      &$2.51 \pm 0.16$             & $4.51 \pm 0.50  $                \\                 
          $a_{P}$  [au]          &$0.77 \pm 0.02$      &$1.60 \pm 0.04$             & $1.82 \pm 0.07  $                \\                 
  \hline

\end{tabular}
\end{center}
~$^{\star}$ The number in parenthesis refer to the model assuming a constant velocity.
\end{table*}
\renewcommand{\arraystretch}{1.}

\section{RV long-term trends and stellar binaries}\label{binaires}

We describe hereafter 14 spectroscopic binaries or massive companions that we identified in our survey using firstly the (RV,~BIS) criterion (Sect.~\ref{classif}). When possible, we fitted the RV data either:
\begin{itemize}
\item with a Keplerian model when the observation time baseline is higher than or covers a significant part of the orbital period, using the {\it yorbit} software. We then removed the binary signal to search for other companions or to characterize stellar activity, and to reach better detection limits.
\item with a first or second-order polynomial fit, when we could notice a clear linear or quadratic trend in the RV over the observation time baseline. We studied the residuals as well to further explore the system.
\end{itemize}
The gains on the RV amplitude and rms obtained after correcting for the binary signal/trend are reported in Table~5. We also looked at the other line profile observables (CCF and FWHM) to further characterize the binary type (SB1, SB2 with large CCF distortions or unresolved SB2) if possible. In addition, we tried to constrain, when possible, the orbital properties of the detected companion: \ie, we derived the possible minimal mass versus sma given the RV trend amplitude and the time baseline. In the case of a RV linear trend $\dot{\nu}$ (equivalent to an acceleration), the companion minimal mass $M_{\rm B}\sin{i}$ is given by $M_{\rm B}\sin{i} = \dot{\nu} a^{2} / G$ \citep{winn09,kane15}, where $a$ is the sma, and assuming a circular orbit. In the case of a quadratic trend, we considered the RV trend amplitude as a lower bound on the RV amplitude that would be induced by the companion (assuming a circular orbit), and then deduced the corresponding minimal mass versus sma relation. We used these constraints to check if the detected RV companion was compatible, as the case may be, with a stellar companion previously detected by direct imaging or astrometric studies. We display the main \harps~data (RV and FWHM time series, (RV,~BIS) diagram, CCF) of our detected binaries in Figs.~\ref{sb1},~\ref{sb2} and~\ref{sb3}, and the constraints on the companion orbital properties in Fig.~\ref{constrain}.

\subsection{\object{HD\,29992}} 
$\beta$ Cae (\object{HIP\,21861}, F3IV-V) shows a long-term quadratic RV trend with a $\sim$1150 \ms~amplitude over the 1762-day ($\sim$4.8 years) time baseline of our observations (Fig.~\ref{sb1}). This trend is best fitted with a $2^{\rm nd}$ order polynomial curve and is associated to a composite (RV,~BIS) diagram, indicative of both the presence of a companion and of high-frequency pulsations. When corrected from the long-term trend, the (RV residuals,~BIS) diagram shows only a remaining vertical spread indicative of pulsations. Taking into account the time baseline, RV trend amplitude and primary mass, and assuming a circular orbit, \object{HD\,29992}B has a minimal mass of $\sim$40 \Mjup, and orbits further away than at least 3 au from its host star (Fig.~\ref{constrain}). It is then either a BD or, more probably, a low-mass star. We classify \object{HD\,29992} as a spectroscopic binary of probable SB1 type, as the target does not show any significant asymmetry in its CCF, nor any significant trend in its FWHM that would be induced by the secondary spectrum (Fig.~\ref{sb1}).

\object{HD\,29992} has not been explicitly identified as a binary before. It was classified as a single star in the multiplicity catalog of \cite{eggleton08}. \cite{ehrenreich10} did not detect any companion to this object, with detection limits excluding companions more massive than 0.07 \Msun~beyond 55 au (Fig.~\ref{constrain}).~$\beta$ Cae has been associated to a \rosat~source by \cite{haakonsen09}, that may be induced by a late-type stellar companion.

\subsection{\object{HD\,49095}} 
\begin{figure*}[ht!]
  \centering
\includegraphics[width=0.85\hsize]{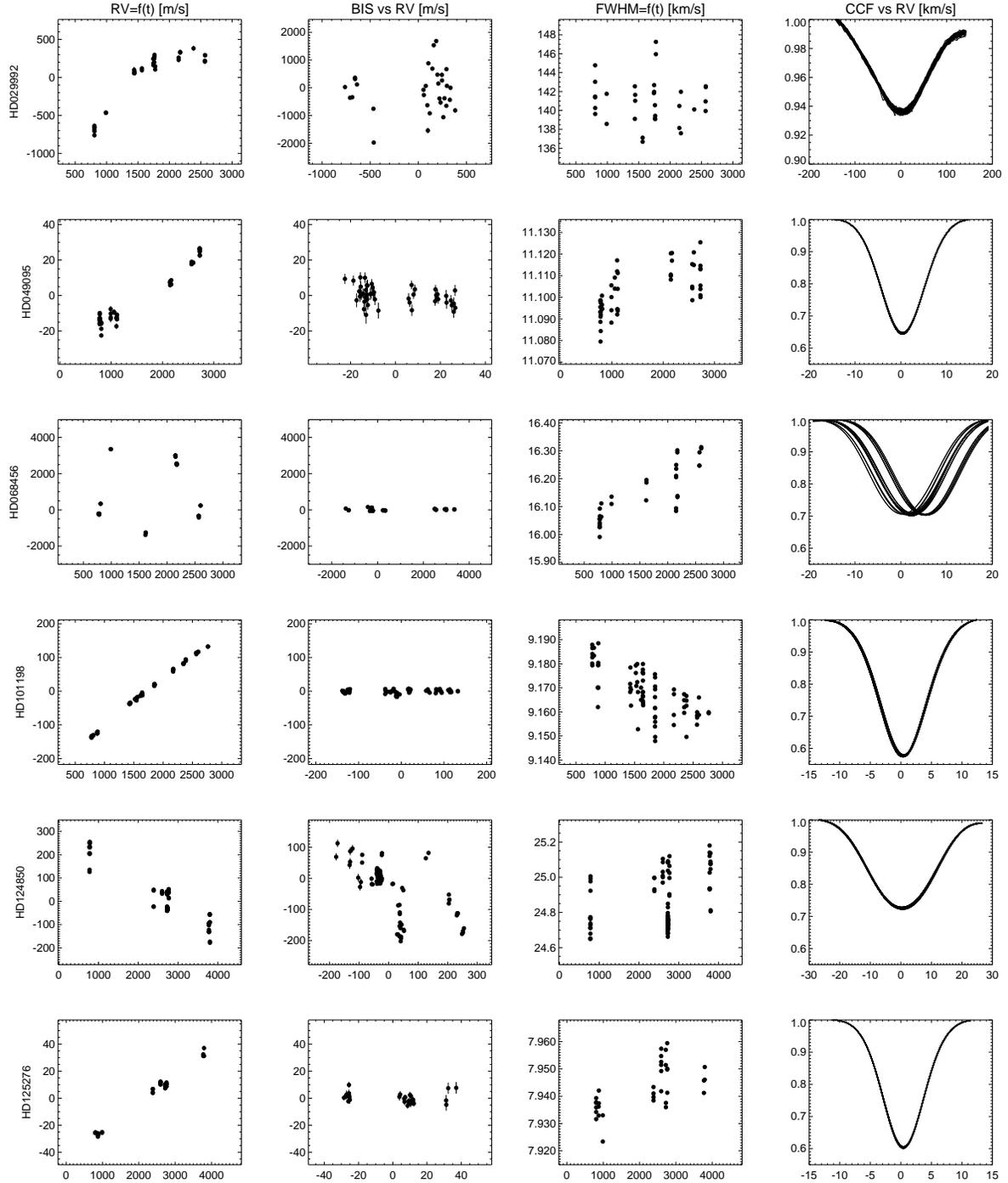}
\caption{Main \harps~data for our detected binaries (first part). {\it From left to right}: RV vs time (Julian Day - 2453000), BIS vs RV, FWHM vs time, stacked CCF. {\it From top to bottom}: \object{HD\,29992}, \object{HD\,49095}, \object{HD\,68456}, \object{HD\,101198}, \object{HD\,124850} and \object{HD\,125276}.}
       \label{sb1}
\end{figure*}

The binarity of this F6.5V target is clear from the long-term quadratic trend in the RV (with a $\sim$49~\ms~amplitude over the 1951-day observation time baseline) and the corresponding flat (RV,~BIS) diagram (Fig.~\ref{sb1}). \object{HD\,49095} FWHM also shows a loose long-term trend with a large short-term dispersion (Fig.~\ref{sb1}), and there is a correlation between the RV and the FWHM data (with a Pearson coefficient of 0.6). We consider that the FWHM long-term variation may be induced by the spectrum of the secondary component of the binary and that \object{HD\,49095} is a possible unresolved SB2 with an orbital period much longer than our observation time baseline. There is no noticeable asymmetry in the CCF (Fig.~\ref{sb1}).

\object{HD\,49095} was flagged as RV variable in \cite{lagrange09}. It was classified as an astrometric binary by \cite{makarov05} and flagged as a proper motion binary in the \hipp~catalog \citep{frankowski07}, although no information is given on the orbital properties of the binary. \cite{ehrenreich10} (hereafter E10) detected a comoving companion which is itself a close (2.3 au) stellar binary with a total mass of 0.11 \Msun~and orbiting at a projected separation of about 31.9 au ($\sim$ 1.3 as) of the primary. Given such parameters and taking \Mstar~$ = 1.2$ \Msun~for the primary \citep{allende99}, the orbital period of the imaged companion is of at least $\sim$160 years (considering in this case that the projected separation from E10 is equivalent to the actual physical separation of the binary components). Although the RV trend we detected could still be induced by a planetary-mass companion (Fig.~\ref{constrain}), it is compatible with the stellar companion imaged by E10. Since the $\sim$1.3 as projected separation derived by E10 is slightly larger than but still close to the \harps~fiber diameter on the sky \citep[1 as][]{pepe02}, the RV-FWHM correlation characteristic of an unresolved SB2 makes it more probable for this imaged companion to be at the origin of the RV trend we detect.

\subsection{\object{HD\,68456}} 
This F6V target shows high-amplitude periodic RV variations associated to a flat (RV,~BIS) diagram (the amplitudes of the RV and of the BIS are of 4739~\ms~and 224~\ms, respectively) and strong CCF variations (but no distortions), illustrative of a SB1 binary with a relatively short period compared to our time baseline (Fig.~\ref{sb1}). We used the {\it yorbit} software to compute the orbital parameters of the companion to \object{HD\,68456}. When letting all the orbital parameters free, we found the best solution to correspond to a stellar companion with a minimal mass of $193 \pm 12$~\Mjup~(or $0.18 \pm 0.01$~\Msun) orbiting at 2.15 au ($\sim$0.1'', corresponding to $P = 898 \pm 0.5$ days) around the primary star on a nearly circular orbit ($e \simeq 0.12$). Once corrected from the keplerian fit, the RV residuals show a very strong anti-correlation with the BIS (Pearson's coefficient $\simeq$-0.96), indicative of stellar activity.

\object{HD\,68456} was first reported as a SB1 by \cite{murdoch93}, who also estimated its orbital parameters. These authors found an orbital period $P \simeq 899 \pm 0.4$ days, an eccentricity $e \simeq 0.12$ and a minimal mass for the secondary of \msini~$= 0.2$ \Msun, in close agreement with our {\it yorbit} best fit. Another fit of the binary orbital parameters was made by \cite{goldin07}, this time on the base of astrometric \hipp~data. These authors fitted the binary orbit with an optimization algorithm and derived a period of $925 \pm 12$ days, an eccentricity of 0.08 and an apparent orbit size of $27.3 \pm 0.8$ mas ($0.58 \pm 0.02$ au). These parameters are also close to our estimation. Remarkably, \cite{goldin07} also provided an estimation of the system inclination, with $i = 30 \pm 5$\degr. By combining our RV minimal mass estimate to this astrometric inclination, we can derive an estimation of the actual companion mass: $386^{+100}_{-70}$~\Mjup~($0.37^{+0.09}_{-0.07}$~\Msun), for such an inclined system. 

\object{HD\,68456} was also observed in direct imaging with NaCo by \cite{ehrenreich10} as part of a search for close companions, but no companion was found. The detection limits derived by E10 exclude companions above 0.07~\Msun~around \object{HD\,68456} beyond 40 au, if accounting for a 2.5 Gyr age as taken from the GCS III. However, a more recent modele atmosphere analysis of the primary \citep{fuhrmann11} argues that \object{HD\,68456} is a probably much older ($\sim$10 Gyr) blue straggler dwarf, and that the secondary is probably a low-mass white dwarf.

\subsection{\object{HD\,101198}} 
We flagged $\iota$ Crt (\object{HIP\,56802}, F6.5V) as a binary based on its RV variations, which are widely dominated by an positive quadratic trend (with a 269 \ms~amplitude over a 1988-day observation time baseline), and its associated flat (RV,~BIS) diagram (Fig.~\ref{sb1}). The FWHM shows a decreasing trend with some dispersion (Fig.~\ref{sb1}), and there is a clear (RV,~FWHM) anti-correlation (Pearson coefficient of -0.7), meaning that the primary spectrum is slightly blended with that of the companion. As for \object{HD\,49095}, we then conclude that \object{HD\,101198} is a probable unresolved SB2 binary with an orbital period much longer than our observation time baseline. The asymmetry in the target CCF is hardly noticeable (Fig.~\ref{sb1}). 

\object{HD\,101198} was flagged as RV variable in \cite{lagrange09}. It is classified as an astrometric binary in the \hipp~catalog \citep{makarov05,frankowski07}. \cite{ehrenreich10} imaged a companion at a projected separation of 25 au ($\sim$0.9 as) with an estimated mass of 0.57 \Msun. Assuming that the 25 au projected separation derived by E10 corresponds to the actual sma and a circular orbit, the minimal orbital period of the imaged companion would be of about 91 years. The drift observed in our RV data is compatible with the properties of the companion imaged by E10 (Fig.~\ref{constrain}), and the $\sim$0.9 as projected separation is compatible with a SB2 (see above).

\subsection{\object{HD\,124850}} 
$\iota$ Vir shows a clear quadratic trend in the RV (with a $\sim$93 \ms~amplitude over the 3024-day baseline) that is best fitted by a $2^{\rm nd}$ order polynomial curve, associated to a composite BIS (RV-BIS correlation coefficient of -0.66, Fig.~\ref{sb1}). Once corrected from the RV binary fit, the residuals show an even stronger anti-correlation with the BIS variations (Pearson's coefficient of -0.83). We conclude that \object{HD\,124850} is a spectroscopic binary and is active.

\object{HD\,124850} was classified as RV variable in \cite{lagrange09}. A candidate stellar companion was reported by \cite{raghavan10}, but no orbital data was available. It has also been associated to a \rosat~source by \cite{haakonsen09}, which points toward a late-type stellar companion. \cite{gontcharov10} derived orbital parameters from a combination of \hipp~data and ground-based astrometric catalogs. The authors estimated an orbital period of 55 years, an apparent photometric sma of $200 \pm 50$ mas (or $4.28 \pm 1.07$ au given the target parallax) and a relative sma of $830 \pm 20$ mas ($17.76 \pm 0.43$ au). They also derived an eccentricity $e$ = $0.1 \pm 0.1$, a mass of $0.6 \pm 0.2$ \Msun~for the secondary (given a 1.53 \Msun~mass for the primary) and a system inclination of $60 \pm 9$\degr. With such orbital and physical parameters, \object{HD\,124850}B would induce a RV semi-amplitude $K$ = 2.4 \kms. For such a RV curve, the possible RV quadratic trends on our observation time baseline are compatible with the observed RV drift (Fig.~\ref{constrain}), and we conclude that the companion we detected with RV is most likely the companion characterized by \cite{gontcharov10}.

\begin{figure*}[ht!]
  \centering
\includegraphics[width=0.85\hsize]{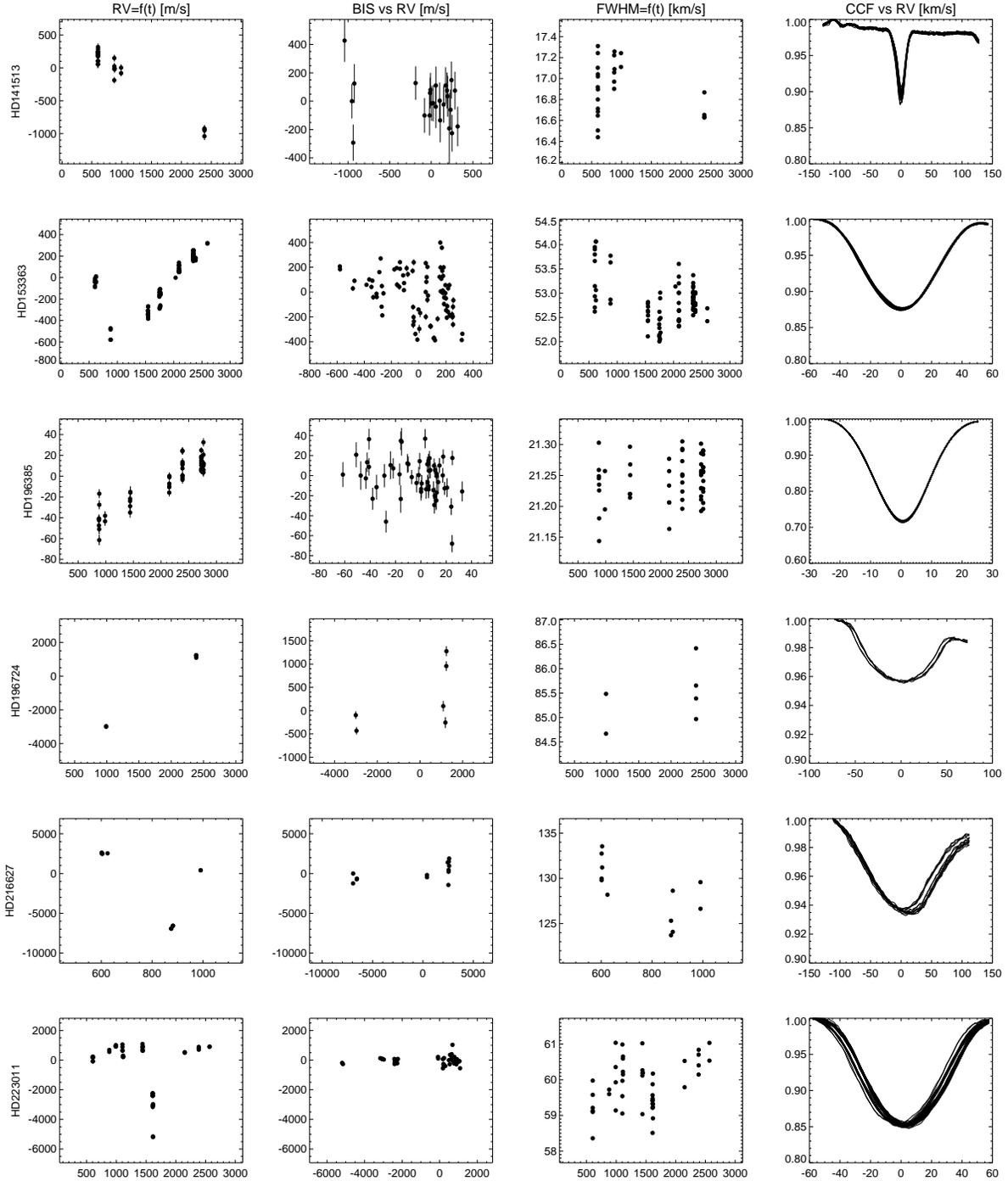}
\caption{Main \harps~data for our detected binaries (second part). {\it From left to right}: RV vs time (Julian Day - 2453000), BIS vs RV, FWHM vs time, stacked CCF. {\it From top to bottom}: \object{HD\,141513}, \object{HD\,153363}, \object{HD\,196385}, \object{HD\,196724}, \object{HD\,216627} and \object{HD\,223011}.}
       \label{sb2}
\end{figure*}

\subsection{\object{HD\,125276}} 
This target shows a strong linear trend in the RV (with a $\sim$65 \ms~amplitude over the 2989-day observation time baseline) and a flat (RV,~BIS) diagram characteristic of a companion (Fig.~\ref{sb1}). It is a representative example of the interest of correcting such long-term trends: the RV rms of the residuals is about ten times smaller than the rms of the original RV data, and its amplitude is nearly eight times smaller. \object{HD\,125276} FWHM shows an increasing trend with time and is correlated to the RV (Pearson's coefficient of 0.65, Fig.~\ref{sb1}), meaning that the system can be an unresolved SB2. The observed RV trend could still be induced by a planetary-mass or stellar companion with an orbital period longer than our time baseline (Fig.~\ref{constrain}).

Based on the comparison of \hipp~and Tycho-2 proper motions, \cite{raghavan10} reported a candidate M-type stellar companion to \object{HD\,125276} with a projected separation of 144 au, but this possible companion has not been retrieved since. There is no other occurrence in the literature of a binary companion to \object{HD\,125276} at our knowledge.

\subsection{\object{HD\,141513}} 
This A0V target shows a linear trend in its RV with a 1358 \ms~amplitude over the 1781-day time baseline, along with a flat (RV,~BIS) diagram, pointing towards a spectroscopic binary status (Fig.~\ref{sb2}). $\mu$ Ser has been reported as an astrometric binary by \cite{makarov05}. Its binary status was also reported and characterized by \cite{malkov12}, who derived a 33.75-year orbital period, a 255 mas (12.2 au) sma and a highly eccentric orbit ({\it e} = 0.75) for the companion. Another set of orbital parameters was derived by \cite{gontcharov10} for $\mu$ Ser, based on astrometric data. The authors reported an apparent photometric sma of $110 \pm 10$ mas ($5.3 \pm 0.5$ au) and a relative sma of $350 \pm 10$ mas ($16.7 \pm 0.5$ au), along with an eccentricity $e = 0.4 \pm 0.3$, a $36 \pm 2$-year orbital period and a mass of $2.4 \pm 0.4$ \Msun~for each of the binary components. Our RV observations are compatible with both the parameters derived by \cite{malkov12} (Fig.~\ref{constrain}) and by \cite{gontcharov10}.

\subsection{\object{HD\,153363}}
 We acquired 85 spectra on {\it 26} Oph (\object{HIP\,83196}, F3V), covering a 1995-day baseline (5.5 years). The RV show large-amplitude ($\sim$901~\ms~on our time baseline) variations that are clearly induced by a companion on an eccentric orbit (Fig.~\ref{sb2}). The associated (RV,~BIS) diagram is composite, indicative of both a companion and of stellar activity (Pearson's coefficient of -0.32). The FWHM shows both short-term and long-term variations, but there is no correlation with the RV (Pearson's coefficient $<$0.1, see Fig.~\ref{sb2}).

We used the {\it yorbit} software to try to fit a Keplerian model to the RV. Given that our time baseline does not cover a complete orbital period of the companion, there is a large uncertainty on the orbital parameters we derived. We consider that the best fit derived with {\it yorbit} gives lower values on the orbital parameters of the companion (especially on the orbital period), rather than a realistic estimation of the parameters. The {\it yorbit} best solution corresponds to a \msini~$= 44$ \Mjup~companion with a $3067.5 \pm 1926.5$-day period (corresponding to a $\sim$4.6 au sma) and an eccentricity $e = 0.31 \pm 0.15$. The residuals of the Keplerian fit are clearly correlated to the BIS (Pearson's correlation coefficient of -0.51), showing that their remaining dispersion is induced by stellar activity.

\begin{figure*}[ht!]
  \centering
\includegraphics[width=0.85\hsize]{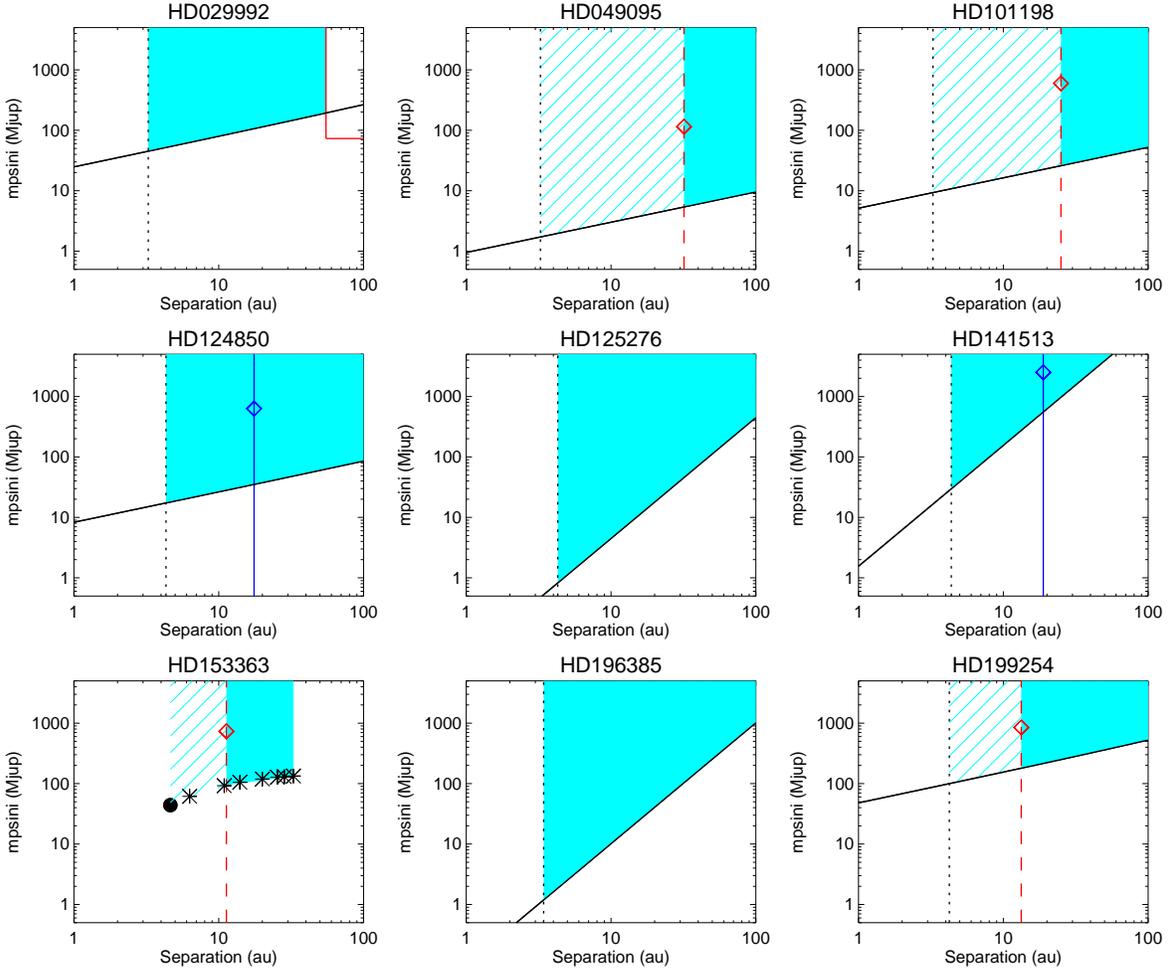}
\caption{Limits on the orbital parameters (\msini, sma) of the companions to our targets with long-term RV trends, given the observed RV variations. {\it Black solid line}: Minimal mass vs sma. {\it Dotted black line}: Minimal sma given the observation time baseline (\ie, considering that the companion orbital period is at least equal to the time baseline, and assuming a circular orbit). In the case of \object{HD\,153363}, the mass and sma corresponding to the Keplerian fit derived with {\it yorbit} with all parameters free is showed as a {\it black filled dot}, and the masses and sma corresponding to the {\it yorbit} fits obtained for longer periods are showed as {\it black crosses}. The area filled and/or shaded in cyan corresponds to the (mass,~sma) domain based on these constraints. {\it Dashed red line}: projected separation of the imaged companion (if any), resulting possible mass from the DI study ({\it red diamond}), or DI detection limits ({\it solid red line}). \object{HD\,29992}, \object{HD\,49095}, \object{HD\,101198}, \object{HD\,153363}: DI from \cite{ehrenreich10}. \object{HD\,199254}: DI from \cite{derosa14}. {\it Blue solid line}: sma of the companion detected through astrometric measurements, and mass estimation ({\it blue diamond}). \object{HD\,124850}: astrometry from \cite{gontcharov10}; \object{HD\,141513}: from \cite{malkov12}.}
\label{constrain}
\end{figure*}

We already flagged this object as a RV variable in \cite{lagrange09}. \object{HD\,153363} was classified as an astrometric binary by \cite{makarov05} and a proper motion binary by \cite{frankowski07}. Finally, \cite{ehrenreich10} imaged a comoving companion to \object{HD\,153363} with a projected separation of 11.3 au ($\sim$0.35 as) and an estimated mass of 0.7 \Msun. No astrometric information is available in the literature, hence we will discuss here the possibility for our detected RV companion to be the object imaged by E10 (hereafter \object{HD\,153363}B). Given the projected separation reported by E10 and taking into account the given mass of 0.7~\Msun, we derived a lower value on \object{HD\,153363}B orbital period of $\sim$9645 days. We then used {\it yorbit} to constrain the RV companion best orbital parameters for larger, increasing orbital periods in the 4800 to 56000-day range (Fig.~\ref{constrain}). For such a period range, the RV companion minimal mass increases from $\sim$62 to $\sim$133 \Mjup, and its eccentricity also increases from 0.42 to 0.85. 

Finally, these results show that the RV variations we detected are compatible with the properties of \object{HD\,153363}B as determined by E10 (Fig.~\ref{constrain}). However, they also show that if the imaged companion corresponds indeed to our RV companion, it should then have a high eccentricity (of at least $\sim$0.6, when assuming that \object{HD\,153363}B projected separation is equivalent to its physical separation to the primary).

\begin{figure*}[ht!]
  \centering
\includegraphics[width=0.65\hsize]{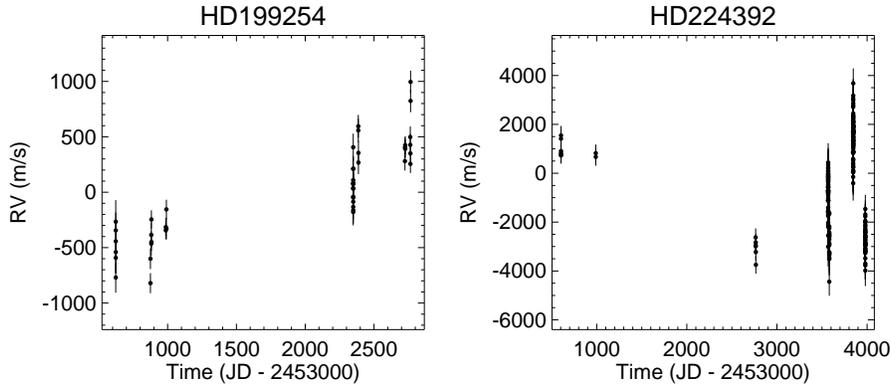}
\caption{\harps~RV for our binaries of very early spectral type (\ie~without line profile computation). {\it From left to right}: RV time series of \object{HD\,199254} and \object{HD\,224392}.}
       \label{sb3}
\end{figure*}

\subsection{\object{HD\,196385}} 
This target shows a linear trend in the RV with a $\sim$94 \ms~amplitude over the time baseline. The (RV,~BIS) diagram is composite (Pearson's coefficient of -0.26, Fig.~\ref{sb2}). Once corrected from the companion-induced trend, the RV residuals have an amplitude twice smaller and a rms twice and a half smaller than before the correction (see Table~\ref{sample}). They also show a significant anti-correlation with the BIS (Pearson's coefficient of -0.45), corresponding to stellar activity. The Lomb-Scargle periodogram of the RV residuals show some peaks at about 1.5 and 3 days that are also present in the BIS periodogram. They are then most probably induced by the conjunction of stellar activity and stellar rotation. \object{HD\,196385} has not been reported as a binary before. Given the RV trend amplitude induced, the companion to \object{HD\,196385} can be either of planetary, BD or stellar nature (Fig.~\ref{constrain}).

\subsection{\object{HD\,196724}} 
We only acquired six spectra on {\it 29} Vul over a 1396-day time baseline. However, they allow us to detect high-amplitude ($\sim$4 \kms) RV variations over the observation time baseline along with BIS and FWHM high-amplitude variations, showing that \object{HD\,196724} is a spectroscopic binary of probable SB2 type. The CCF are also clearly variable (Fig.~\ref{sb2}), however the very small spectrum number prevent us from deriving any constraints on the properties of \object{HD\,196724AB}. {\it 29} Vul is flagged as a proper-motion binary by \cite{frankowski07} and \cite{makarov05}. 

 \subsection{\object{HD\,199254}}
With too few spectral lines to compute the line profiles, we rely only on the RV data to classify this A5V target as a spectroscopic binary. The 45 RV measurements show a $\sim$2000 \ms-amplitude quadratic trend on the 2143-day observation time baseline (Fig.~\ref{sb3}), pointing towards a stellar nature (Fig.~\ref{constrain}). \cite{derosa14} reported the detection of a stellar companion to \object{HD\,199254} through direct imaging, at a projected separation of 13.3 au and with an estimated mass of 0.81 \Msun. Such a companion is compatible with the RV trend we detected.

\subsection{\object{HD\,216627}} 
We acquired only 13 spectra on $\delta$ Aqr (\object{HD\,216627}, \object{HIP\,113136}), covering a 389-day time baseline. High-amplitude ($\sim$9.5 \kms) RV variations, a flat (RV,~BIS) diagram, a clear asymmetry in the CCF (with a broader RV span on their red wing than on their blue wing), and corresponding FWHM variations (that are correlated to the RV with a Pearson's coefficient of 0.79) point unambiguously towards a SB2 status for this A3V target (Fig.~\ref{sb2}).

 $\delta$ Aqr was already reported as a binary in \cite{lagrange09}. A possible member of the Ursa Majoris (UMA) moving group \citep[according to][]{king03}, it was first reported as an astrometric binary by \cite{goldin07} on the base of \hipp~data. These authors derived a first set of orbital parameters, with a $483 \pm 20$-day period, a $0.12 \pm 0.2$ eccentricity and an inclination of $41 \pm 16 \degr$ from edge-on. \cite{ehrenreich10} reported no detection in deep imaging, excluding companions more massive than 0.07 \Msun~beyond 100 au. Yet very interestingly, the companion to \object{HD\,216627} was directly detected through IR interferometry made at the Very Large Telescope Interferometer (VLTI) \citep{absil11}. The authors used the \pionier~four-telescope interferometer to obtain precise closure phase measurements that allowed them to unambiguously detect the companion. These authors derived a contrast of 4.2 in the $H$-band, leading them to estimate the companion spectral type to be around G5V. However, due to the poor coverage of the ({\it u,v}) plan, they were not able to constrain the separation between the two components of the binary, giving only three preferred positions at 37.4, 41 and 46.5 mas from the primary (\ie~projected separations of 1.9, 2 and 2.3 au respectively).

The temporal sampling of our own RV measurements is not sufficient to constrain the orbital parameters of $\delta$ Aqr with a Keplerian fit. Periods can be fitted down to less than 10 days (corresponding to minimum masses of a few tens of Jupiter masses), but periods in the range 300-700 days can equally be fitted. The latter period range is compatible with the period derived by \cite{goldin07} and the interferometric separations derived by \cite{absil11}.

\subsection{\object{HD\,223011}}

\object{HD\,223011} (\object{HIP\,117219}, A7III-IV) shows high-amplitude RV variations ($\sim$6.3~\kms~peak-to-peak) and a flat (RV,~BIS) diagram that clearly show the presence of a binary companion on an eccentric orbit. The CCF data shows strong shifts in radial velocity characteristic of a SB1 but no noticeable asymmetry that would point towards a SB2 status (Fig.~\ref{sb2}). We acquired 46 spectra over 1936 days on \object{HD\,223011}. Yet, the temporal sampling of our observations is not complete enough to fully constrain the binary orbital parameters: even if our time baseline may cover several orbital periods, it happens that most of our data points seem to be located in a restricted phase range, not fully covering the RV variations induced by the companion. When looking at the RV Lomb-Scargle periodogram, multiple peaks are still present above the $1\%$ FAP, with the most pre-eminent at periodicities of about 114, 140, 37 and 92 days. We used the {\it yorbit} software to fit the RV with a single Keplerian model. Without putting any prior constraint on the orbital parameters, the best solution corresponds to a companion with a minimal mass of 69~\Mjup~on an eccentric orbit ($e = 0.62$) with a 37.7-day period (sma $=$ 0.27 au). As in the case of \object{HD\,153363}, we consider that this fit gives a lower value on the binary orbital parameters but not a definitive determination. The residuals of the fit show a 111.2 \ms~dispersion without any remaining significant periodicity in their periodogram.

We tried to fit \object{HD\,223011} RV data with single Keplerian models corresponding to longer ranges of orbital periods (up to 3000 days) by setting different prior constrains on the period with {\it yorbit}. However, all solutions exhibit very high eccentricities (above 0.8) and a significantly larger dispersion of the residuals (always above 200 \ms). Given that there is no significant asymmetry in the CCF (characteristic of a SB2), the close companion to \object{HD\,223011} has probably a contrast of at least 2-3 magnitudes in the $V$-band with respect to the primary, as a brighter companion would significantly impact the \harps~spectra and distort the CCF. We thus conclude that \object{HD\,223011} has a close low-mass stellar companion, with a minimum period of 37.7 days on an eccentric orbit. \object{HD\,223011} was already classified as RV variable in \cite{lagrange09}, and \cite{ehrenreich10} did not detect any companion with deep imaging. The detection limits derived by E10 exclude companions more massive than 0.07 \Msun~beyond 130 au.

\subsection{HD\,224392}
Given its high \vsini~(190 \kms), we do not have BIS data for $\eta$ Tuc (\object{HIP\,118121}, A1V). The RV show high-amplitude variations ($\sim$8 \kms), much larger than the RV dispersion (rms $\sim$1.96 \kms) induced by the fast rotation (Fig.~\ref{sb3}). We conclude that \object{HD\,224392} is a probable spectroscopic binary. A stellar companion to $\eta$ Tuc \citep[member of the $\sim$30 Myr Tucana Horologium young association, see][]{zuckerman11} was directly detected thanks to closure phase and squared visibility measurements made with the \pionier~IR interferometer on the VLTI \citep{marion14}. These authors stressed nonetheless that the detection remained somewhat suspicious as the detected companion position was not the same depending on the data type (closure phase or square visibility measurements). New \pionier~observations seem to confirm however the existence of the stellar companion \object{HD\,224392B}; these new data as well as the properties of this companion will be analyzed in more details in a forthcoming paper.

\section{Stellar intrinsic variability}\label{intr_var}

\begin{figure*}[ht!]
  \centering
\includegraphics[width=0.85\hsize]{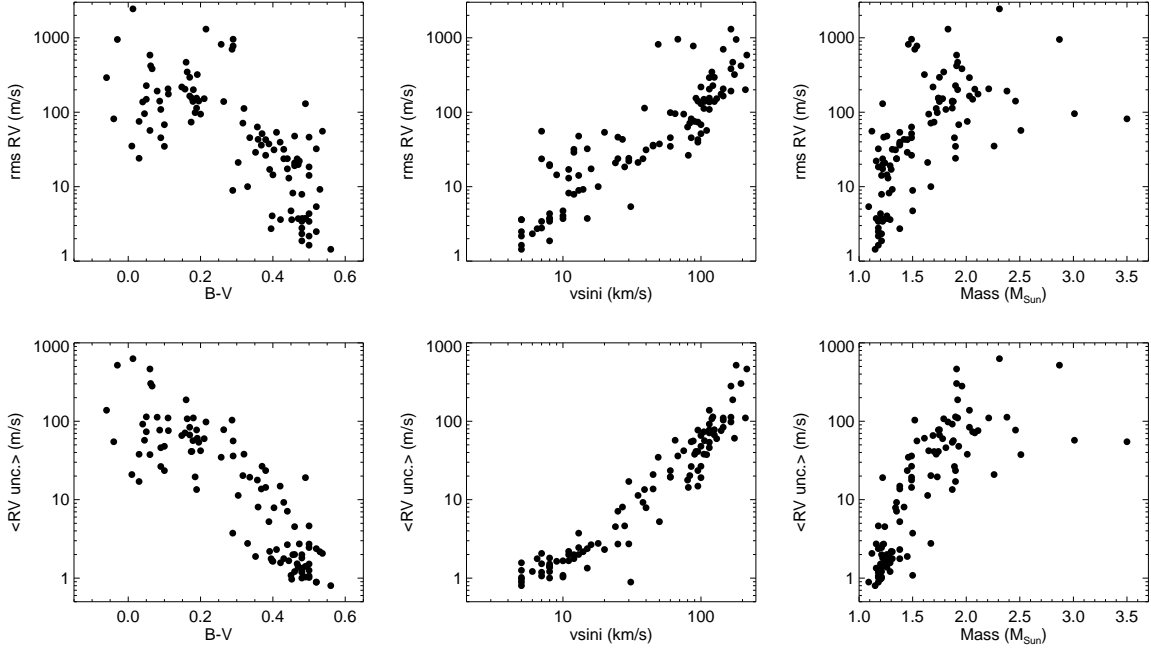}
\caption{Stellar intrinsic RV variability vs stellar properties. For each of our targets, we display the RV rms ({\it first row}) and the averaged RV uncertainty ({\it second row}) vs \bv, \vsini~and \Mstar~({\it from left to right}).}
       \label{stell_var}
\end{figure*}

\begin{figure*}[ht!]
  \centering
\includegraphics[width=0.85\hsize]{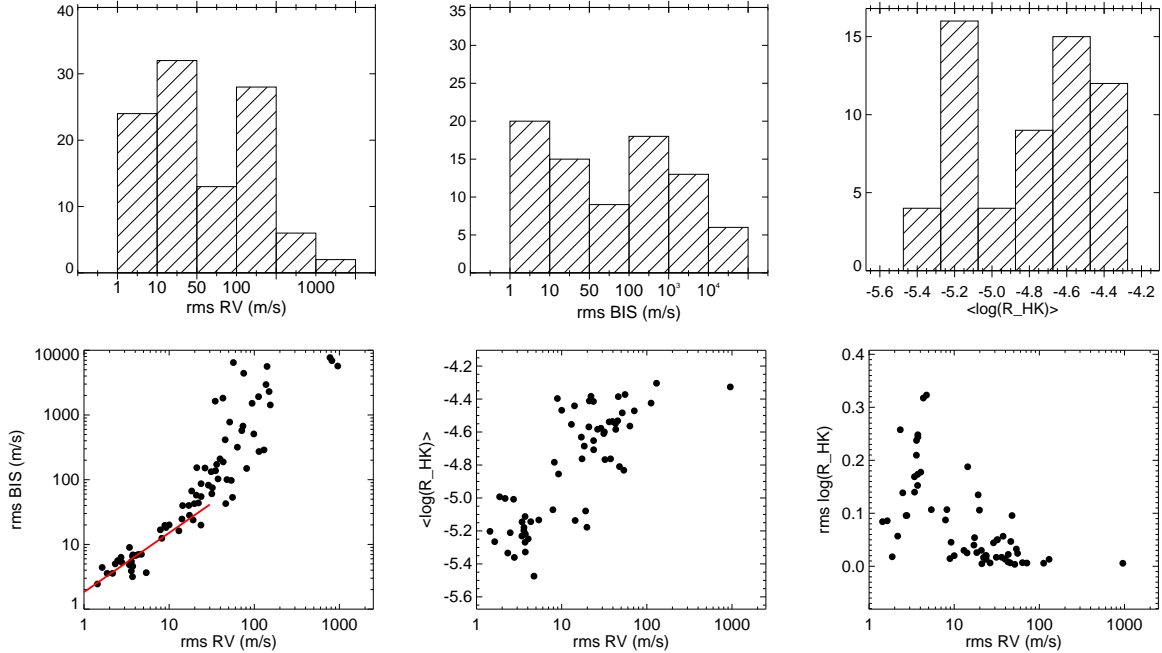}
\caption{Stellar intrinsic variability in our observables. {\it First row}: RV rms histogram; BIS rms histogram; mean \rhk~histogram. {\it Second row}: BIS rms vs RV rms; mean \rhk~vs RV rms; \rhk~rms vs RV rms.}
       \label{hist_var}
\end{figure*}

Stellar magnetic activity (spots and faculae) and high-frequency pulsations are the two main sources of intrinsic RV variability for our targets \citep[detailed examples can be found in][]{lagrange09}. We display in Fig.~\ref{stell_var} the RV rms and the mean RV uncertainty versus our target physical properties (\bv, \vsini, stellar mass), after substracting companion-induced RV variations. As expected, the RV rms and the RV mean uncertainty are strongly correlated with the spectral type and the \vsini~of our targets; stars of earlier ST (with fewer absorption lines that are more broadened by the rotation rate) have higher RV uncertainties. Earlier-type, faster-rotating stars being also the most massive ones, the RV dispersion and mean uncertainty increase with an increasing \Mstar. If taken as a whole, the RV rms is larger by an order of magnitude than the typical RV rms for FGK dwarf surveys \citep[\eg~][]{howard11}. The ``two-peak'' shape of the RV and BIS rms distributions (Fig.~\ref{hist_var}) is probably explained by the relative lack of late A/early F dwarfs in our sample (Sect.~\ref{data}).\\

We display in Fig.~\ref{hist_var} the BIS rms versus RV rms distribution. This distribution clearly shows two regimes: {\it i}) for RV and BIS rms below $\sim$30 \ms, the BIS rms increases linearly with the RV rms (Fig.~\ref{hist_var}); {\it ii}) for higher rms, the BIS rms increases much fastlier than the RV rms. These two regimes correspond to the two main sources of RV jitter in our sample, respectively stellar magnetic activity and high-frequency pulsations. We also display in Fig.~\ref{hist_var} the mean \rhk~and \rhk~rms versus RV rms distributions. Here, the mean \rhk~increases steadily with the RV rms, which is in agreement with it being a proxy for the stellar magnetic activity level. On the contrary, the \rhk~dispersion decreases as the RV rms increases.\\

For some of our later-type, active stars, we can at least partially correct the RV data from the activity-induced jitter by using the correlation between BIS or \rhk~and RV activity-induced variations. We considered here to correct the RV from activity only for our targets with a clear enough (RV,~BIS) or (RV,~\rhk) correlation. We used the Pearson correlation coefficient to define a quality criterion for the correlation and decided that a minimal absolute Pearson coefficient of 0.7 would be required to consider the (RV,~BIS) or (RV,~\rhk) correlation significant enough. We found that eleven of our targets met this criterion (two of them having already been corrected from a binary trend) in the case of the (RV,~BIS) correlation, and only one in the case of the (RV,~\rhk) correlation (note that for this target, \object{HD\,25457}, we both correct the activity jitter on the short-term with the RV-BIS correlation and on the long-term with the RV-\rhk~correlation). For these targets, we applied a linear fit to the (RV,~BIS) or (RV,~\rhk) data and then removed it from the RV. We display the list of these targets, along with the gains in RV amplitude, rms and on the detection limits in Table~\ref{tab_bis_corr}. 

\begin{table*}[ht!]
\caption{Correction from activity-induced RV jitter}
\label{tab_bis_corr}
\begin{center}
\begin{tabular}{c c c c c c c c}\\
\hline
\hline
HD     & Correction          & RV correlation & RMS ratio~$^{\star}$ & Ampl. ratio~$^{\star}$ & 10-day limdet.           & 100-day limdet.   & 1000-day limdet \\
       &   type              & Pearson coef.  &                    &                       & ratio~$^{\star}$          &  ratio~$^{\star}$   & ratio~$^{\star}$           \\
       &   (BIS or \rhk) & (absolute)     &                    &                       &                         &                    &                     \\
\hline
25457  & BIS                 & 0.71           & 0.71               & 0.78                  & 0.61                    & 0.78                      & 0.79     \\
 -     & \rhk~($\dagger \dagger$)& 0.82           & 0.42               & 0.51                  & 0.28                    & 0.21                      & 0.05    \\
30652  & BIS                 & 0.79           & 0.62               & 0.68                  & 0.52                    & 0.47                      & 0.42 \\
31746  & BIS                 & 0.81           & 0.58               & 0.58                  & 0.80                    & 0.81                      & 1.00 \\
33262  & BIS                 & 0.82           & 0.58               & 0.59                  & 0.65                    & 0.42                      & 0.60\\
49933  & BIS                 & 0.81           & 0.58               & 0.44                  & 0.67                    & 0.62                      & 0.92\\
68456 ($\dagger$) &  BIS     & 0.96           & 0.29               & 0.34                  & 0.23                    & 1.                        & -  \\
76653  &  BIS                & 0.73           & 0.68               & 0.70                  & 0.90                    & 0.74                      & 0.30 \\
114642 &     BIS             & 0.89           & 0.46               & 0.42                  & 0.52                    & 0.82                      & 0.75\\
124850 ($\dagger$) &   BIS   & 0.83           & 0.52               & 0.63                  & 0.48                    & 0.77                      & -    \\
138763 &              BIS    & 0.96           & 0.27               & 0.30                  & 0.28                    & 0.39                      & 0.36\\
219482 &             BIS     & 0.83           & 0.52               & 0.59                  & 0.57                    & 0.51                      & 0.67\\
\hline
\end{tabular}
\end{center}
~$\dagger$ After correction from binary trend.\\
~$\dagger \dagger$ After first removing the (RV,~BIS) correlation.\\
~$^{\star}$ Ratio of the parameter values taken after to before activity correction. In the case of \object{HD\,25457}, two corrections are performed successively; for both corrections we display the ratio values related to the RV taken without correction.
\end{table*}

\section{Detection limits}\label{limdets}

\subsection{Detection limit determination}

We estimated the detection limits for each target of our survey, \ie~the upper limit on the \msini~of an hypothetic companion that would be detected given the observed RV, at different orbital periods. We computed the detection limits on a grid of 200 log-spaced orbital periods in the 1 to 1000-day range, roughly corresponding to sma in the 0.02 to 2.5-au range. We used the Local Power Analysis (hereafter LPA) method developed by \cite{meunier12}. These detection limits are computed assuming circular orbits. The LPA method compares the maximum power $Pw_{\rm pl}$ of the RV signal induced by a fake planet of a given mass and a given period (with the same temporal sampling as the actual data) to $X$ times the maximum power $Pw_{\rm dat}$ of the actual RV signal within a localized period range of the periodogram (0.75P-1.25P for a given period P). Contrary to \cite{meunier12}, who took $X = 1$, we used here a slightly more restricting $X = 1.3$ ratio, empirically chosen to correspond to the ``naked-eye'' detection of a real planet signal in an observed periodogram. The fake planet \msini~is considered to be above the detection limit if for 100 phase realizations (\ie~a 99\%~confidence level), $Pw_{\rm pl}$ is always above $ X . Pw_{\rm dat}$. For each orbital period of the grid, the fake planet decreasing \msini~are tested iteratively during several loops, with a narrower mass step at each loop. The upper limit on the considered \msini~range is 100 \Mjup, and the finest mass step (equivalent to the step of the \msini~grid) is 0.005 \Mjup. The LPA method gives lower detection limits than bootstrap methods for various stellar types and RV jitter levels \citep[][]{meunier12,lagrange13}. When the target was found to have a companion or to be active, we compute the detection limits before and after removing the linear/quadratic RV trends, keplerian fits or stellar activity correlations.

Note: assuming circular orbits only to compute the detection limits is the common approach in the literature of large RV surveys \citep[see \eg][]{cumming08,lagrange09,howard11}, as: {\it i}) it allows to considerably reduce the computation time and {\it ii}) the eccentricity impact on planet detectability is relatively small for eccentricities up to $\sim$0.5 \citep[see][and ref. therein]{cumming08}. However, for high eccentricities, the planet detectability can be highly affected as it will increasingly depend on the phase coverage of the planet orbit. Thus, considering one of our target among others, including eccentric orbits in our upper limit computation would make our mass detection limits for this target higher at specific periods. When considering the whole survey, our mass detection limits would be slightly higher.

\begin{figure*}[ht!]
\centering
\includegraphics[width=0.85\hsize]{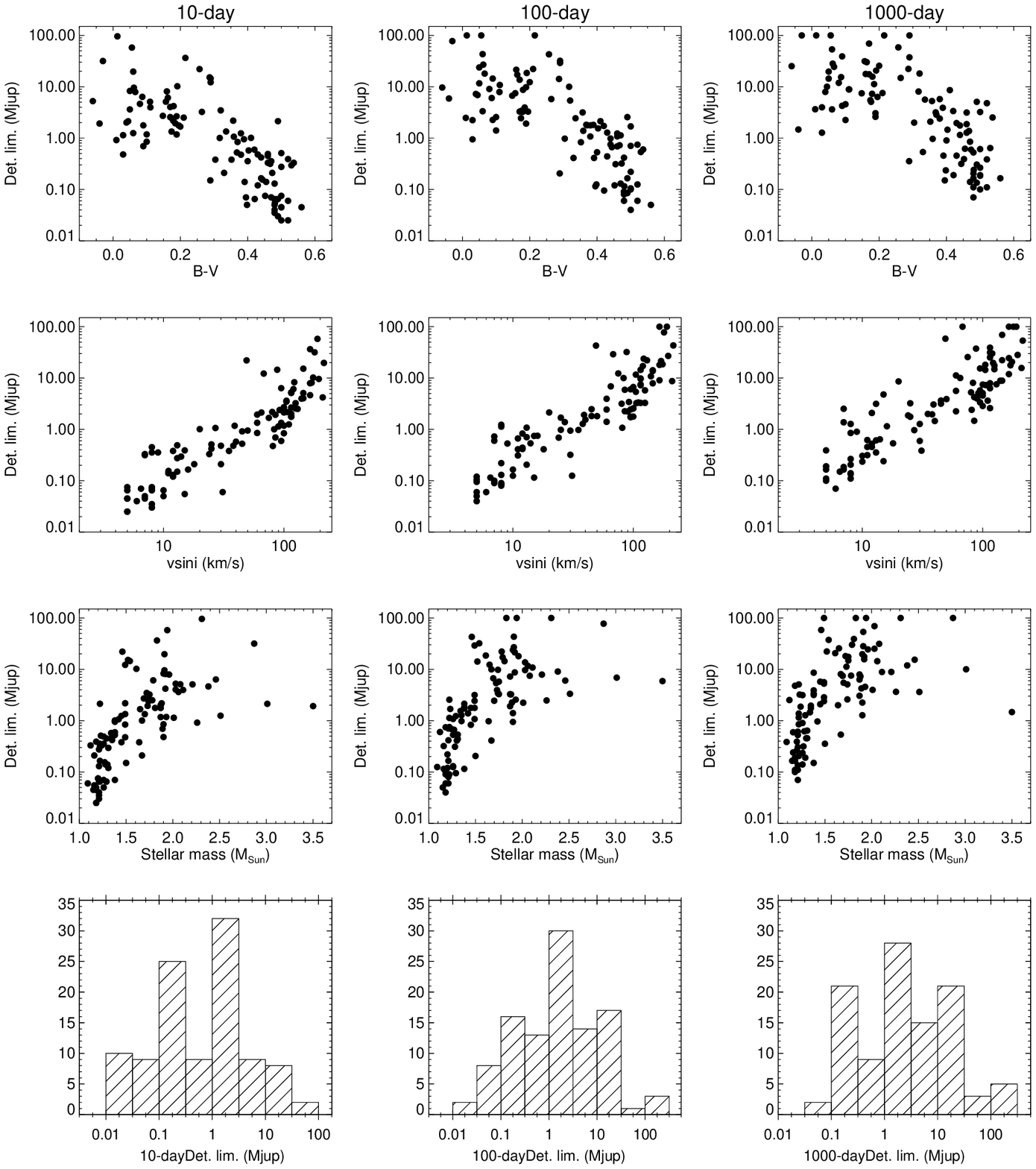}
\caption{Achieved detection limits for our sample versus main stellar parameters. For all panels, each black dot represents the detection limit for one target. All targets are displayed except for stars with detected GP companions. {\it Left}: mean detection limit in the range 1-10 days; {\it middle}: the same in the range 10-100 days; {\it right}: the same in the range 100-1000 days. {\it Top}: detection limits vs \bv; {\it middle top}: vs \vsini; {\it middle bottom}: vs \Mstar; {\it bottom}: in histograms.}
\label{lim_vs_param}
\end{figure*}

\subsection{Results}

\subsubsection{Detection limits versus physical properties}
We display in Fig.~\ref{lim_vs_param} the detection limits achieved with the LPA method versus our sample main parameters (\bv, \vsini~and \Mstar). Our LPA detection limits are clearly correlated to these parameters, with higher detection limits for increasing \vsini~and for earlier ST. This is in agreement with our results in terms of stellar variability (Sect.~\ref{intr_var}). In terms of \vsini, we can roughly distinguish two populations:
\begin{itemize}
\item for \vsini~lower than 100 \kms, the detection limits are mostly in the planetary domain ($\lesssim$10-20 \Mjup), going down well below 1 \Mjup~down to $\sim$0.1 \Mjup~at all periods in the best cases;
\item for \vsini~greater than 100 \kms, the detection limits are mostly larger than 1 \Mjup; still a large number of them remain below 10 \Mjup at all periods, while the remainder are in the brown dwarf domain.
\end{itemize}
In terms of spectral types, we can also make a rough distinction between our earlier (\bv~$\lesssim$ 0.3) and later type targets:
\begin{itemize}
\item the detection limits achieved for the late-type stars are in the planetary domain, even for periods up to 1000 days;
\item for the earlier-type targets, the detection limits spread over the GP and BD mass domains, but can still go down to a fraction of \Mjup~for periods of a few to a few tens days, and down to a few \Mjup~for periods up to 1000 days.
\end{itemize}
In terms of stellar masses, we can roughly make the same distinction between our targets with \Mstar~$\leq$1.5 \Msun, for which the detection limits remain in the planetary domain for periods up to 1000 days, and our targets with higher masses, for which the detection limits spread from less than 1 \Mjup~at short periods at best to 100 \Mjup~at worst. Finally, we note that if extrapolating these detection limits to the solar parameter values, we would find masses in the range 10-30 \ME, \ie~in the same range than the detection limits we deduced for the Sun based on simulated RV time series of solar magnetic activity \citep[][]{meunier13}.\\

When compared to the typical detection limits obtained by \eg~\cite{cumming08} or \cite{howard11} for later-type (FGK) stars in similar period ranges, we find our detection limits to be generally higher by one order of magnitude in minimal mass. We note that:
\begin{itemize}
\item for very short periods ($\lesssim$ 50 days), 90\%~of our targets have detection limits in the planetary domain, meaning that ``Hot Jupiters'' can be detected by RV for very early ST and for very high rotation rates (up to a hundred of \kms);
\item the same percentage of our targets (90\%) have detection limits lower than $\sim$80 \Mjup~for periods up to 1000 days;
\item 70\%~of our targets have detection limits lower than 10-20 \Mjup~for periods up to 1000 days, meaning that close-in GP (\ie~located at separations similar to most of the GP found by RV around solar-like stars) are detectable around most of the AF MS dwarfs;
\item 30\%~of our targets have detection limits lower than 1 \Mjup~for periods up to 1000 days, meaning that Saturn-like and even Neptune-like planets can be detected around the most favourable of our targets, \ie~mid to late-F dwarfs with low to intermediate \vsini~(up to 20-30 \kms).
\end{itemize}
Contrary to what was commonly expected, these results confirm that GP are widely detectable around Main-Sequence dwarfs of early spectral type, and prove the usefulness of extending RV surveys from FGK-type stars to AF-type dwarfs.

\subsubsection{Sample search completeness}

\begin{figure}[ht!]
\centering
\includegraphics[width=0.95\hsize]{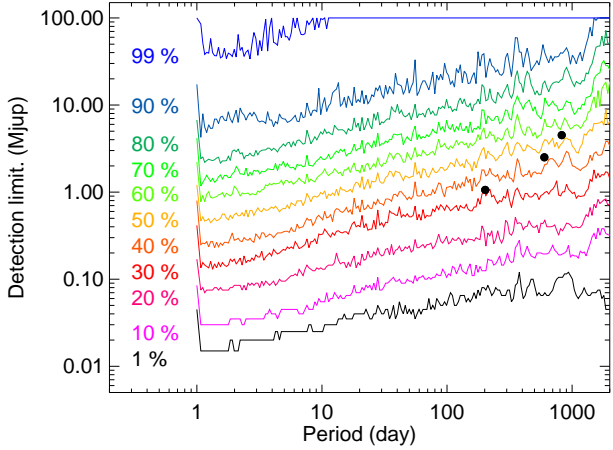}
\caption{GP search completeness of our survey. {\it Black dots}: detected GP (\msini~vs. $P$). {\it Solid lines}: survey search completeness (\ie, fraction of stars with good enough detection limits to rule out planets of a given \msini~at a given orbital period $P$). {\it From bottom to top}: 1\%, 10\% to 90\% (10\% step) and 99\% search completenesses.}
\label{completness}
\end{figure}

To better characterize our results in terms of detection limits, we derived the search completeness function $C$($P$, \msini) of our sample in a way similar to \cite{howard11}, who derived occurrence rates of low-mass planets with short orbital periods around FGK solar-like stars. For given $P$ and \msini, we have:
\begin{equation}\label{Ccomp}
C = \frac{1}{N} \displaystyle\sum_{i=1}^{N} \delta_{i}
\end{equation}
where for each target $i$ of our sample without a detected planet, $\delta_{i} = 1$ if \msini~is higher than the target detection limit at the period $P$ and $\delta_{i} = 0$ otherwise; $N$ being the number of such targets in our sample. The search completeness $C$ therefore gives the fraction of stars of our sample for which we can rule out a ($P$, \msini) planet. Here, we compute $C$(P, \msini) over all our targets for which we recorded at least $N_{\rm m} = 15$ spectra (excluding two targets for which we had too few spectra: \object{HD\,196724} and \object{HD\,216627}). The completeness must be computed only on stars with no detected planets, therefore we had $N = 104$ targets taken into account in Eq.~\ref{Ccomp} (excluding two stars with detected GP: \object{HD\,60532} and \object{HD\,111998}). We took the detection limits into account in the computation of $C$(P, \msini) once we corrected the RV from binary trends or stellar activity correlations, if any. We display $C$($P$, \msini) in Fig.~\ref{completness}.\\

\subsection{GP occurrence rates}

Taking into account the GP detections and detection limits of our sample, we derived a first estimation of the close-in GP occurrence rate around AF, MS stars. The method we followed to deduce GP occurrence rates, detailed in Sect.~\ref{occ_method}, is largely inspired from \cite{howard11}. These authors counted the number of systems $n_{\rm det}$ with detected planets in five given domains of P, \msini. Then, they derived an estimation of the number of planets potentially missed $n_{\rm miss}$ in each of the ($P$,~\msini) domains, based on the number of actual detections and on the survey completeness in these domains. They finally computed the planet occurrence rates taking into account both $n_{\rm det}$ and $n_{\rm miss}$ and using binomial statistics.

\subsubsection{Method}\label{occ_method}

We detail here step by step the method we used to derive our GP detection rates. 
\paragraph{[Mass; period] ranges --} We consider three \msini~domains: {\it i}) a 13 to 80-\Mjup~range, corresponding to BD companions; {\it ii}) a 1 to 13-\Mjup~range, corresponding to Jupiter-like or super-Jupiter GP; and {\it iii}) a 0.3 to 1-\Mjup~range, corresponding to Saturn-like or sub-Jupiter GP. We consider this partition as the most meaningful since it fits well our results and since dividing our achieved \msini~range into more domains would not lead to more significant results, due to small-number statistics. We then consider five different orbital period ranges: {\it i}) the full 1 to 1000-day range over which we computed our LPA detection limits, to derive occurrence rates for all close-in GP; {\it ii}) a 1 to 10-day and a 1 to 100-day ranges to roughly cover the ``Hot Jupiter'' space and allow a comparison with \eg~\cite{cumming08} and \cite{howard11}; {\it iii}) the complementary 10 to 100 and 100 to 1000-day ranges. 

\paragraph{Search completeness function --} Considering a ($P$,~\msini) domain {\it D} with \msini~in the range $m_{\rm p1}\sin{i}$-$m_{\rm p2}\sin{i}$ and $P$ in the range $P_{1}$-$P_{2}$, we compute the search completeness function $C$($P$, \msini) over {\it D} as:

\renewcommand{\arraystretch}{1.25}
\begin{equation}
C_{D} = \frac{\displaystyle\sum_{P_{1}}^{P_{2}} \displaystyle\sum_{m_{\rm p1}\sin{i}}^{m_{\rm p2}\sin{i}} (\frac{1}{N} \displaystyle\sum_{i=1}^{N} \delta_{i}) \ \mathrm{d}P \ \mathrm{d}(m_{\rm p}\sin{i})}{\displaystyle\sum_{P_{1}}^{P_{2}} \displaystyle\sum_{m_{\rm p1}\sin{i}}^{m_{\rm p2}\sin{i}} 1 \ \mathrm{d}P \ \mathrm{d}(m_{\rm p}\sin{i})}
\end{equation}
\renewcommand{\arraystretch}{1.}
with $N = 104$ targets. We compute $C_{D}$ on the same period grid we used to compute the LPA detection limits, and on a mass grid fine enough (20000 mass bins from 0 to 100 \Mjup) to match the mass accuracy reached in the final LPA loop.

\begin{figure}[ht!]
\centering
\includegraphics[width=0.95\hsize]{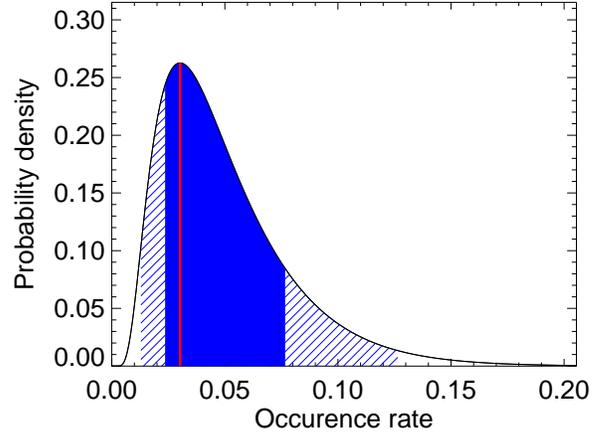}
\caption{Probability density function of the GP occurrence rate for planetary \msini~in the range 1-13 \Mjup~and for periods in the range 1-1000 days, after correcting for the missed GP. {\it Red}: Most probable value. {\it Blue}: 1-$\sigma$ confidence interval. {\it Hatched blue}: 2-$\sigma$ confidence interval.}
\label{binomial}
\end{figure}

\paragraph{``Missed planets'' estimation --} To derive significant occurrence rates, we have to correct our results (in terms of actual detections) from our search incompleteness, \ie~by statistically estimating the number of GP $n_{\rm miss}$ that we potentially missed in our survey. We compute $n_{\rm miss}$ in the same way as in \cite{howard11}. For each ($P$,~\msini) domain with a search completeness $C_{D}$ and with $n_{\rm det}$ systems having one or several detected GP, we have (provided that $n_{\rm det} \geq 1$):
\begin{equation}
n_{\rm miss} = n_{\rm det} (\frac{1}{C_{D}} -1)
\end{equation}
Compared to \cite{howard11}, our sample is much smaller both in terms of targets and of detections. We have therefore several ($P$,~\msini) domains with no actual detection and with a search completeness well below 100\%. In such a case ($n_{\rm det} = 0$ and $C_{D} < 100$\%), we use a number $n_{\rm det}^{'} = 1$ instead of $n_{\rm det} = 0$, and compute $n_{\rm miss}^{'} = n_{\rm det}^{'} (\frac{1}{C_{D}} -1) = \frac{1}{C_{D}} - 1$. We then use these $n_{\rm det}^{'}$ and $n_{\rm miss}^{'}$ in the same way as $n_{\rm det}$ and $n_{\rm miss}$ to compute the companion occurrence rate. Such an approach may appear conservative but we consider it to be the most secure given our target sample size.

\paragraph{Occurrence rate computation --} For each of our ($P$,~\msini) domains, we finally derive the GP occurrence rate and its 1 and 2$\sigma$ uncertainties in the same way as \cite{howard11}, \ie~by using binomial statistics:
\begin{itemize}
\item when $n_{\rm det} \geq 1$, we compute the binomial distribution of $n_{\rm det}$ systems with at least one GP among the $N_{\star} = 106$ targets with 15 or more spectra of our sample. We then multiply the probability distribution by ($n_{\rm det}$+$n_{\rm miss}$)/$n_{\rm det}$ to account for the missed GP, and thus obtain the probability density function (hereafter PDF) of the GP occurrence rate\footnotemark. The GP occurrence rate corresponds to the probability $f$ for which the PDF takes its maximum value (Fig.~\ref{binomial}). To derive the 1 and 2$\sigma$ confidence intervals, we integrate the PDF and compute the $1$ and $2\sigma$ standard deviations. As an example, we display in Fig.~\ref{binomial} the 1 to 13-\Mjup~GP occurrence rate PDF for the full period range of our AF survey, after correcting for the missed GP;
\footnotetext{If $n_{\rm det} \geq 1$, the probability $f$ of drawing $n_{\rm det}$ systems with detected GP among the $N_{\star}$ targets of our sample is $f_{n_{\rm det};N_{\star}}(p) = \left(\!
  \begin{array}{c}
    N_{\star} \\
    n_{\rm det}
  \end{array}
  \!\right) . p^{n_{\rm det}} (1-p)^{N_{\star}-n_{\rm det}}$ with $p$ the probability of having at least one GP around one star.}
\item when $n_{\rm det} = 0$, we compute the binomial distribution of $n_{\rm det}^{'} = 1$ system with GP among the $N_{\star}$ targets and multiply the probability distribution by $1 + n_{\rm miss}^{'}$. Then, we derive the companion occurrence rate in the same way as above, \ie~by taking the probability for which the PDF is maximal. In this case, we consider the lower limit of the confidence intervals to be 0, and we compute the upper limit of the confidence intervals in the same way as in the $n_{\rm det} \geq 1$ case. 
\end{itemize}

\subsubsection{Results and discussion}

\begin{figure}[ht!]
\centering
\includegraphics[width=0.85\hsize]{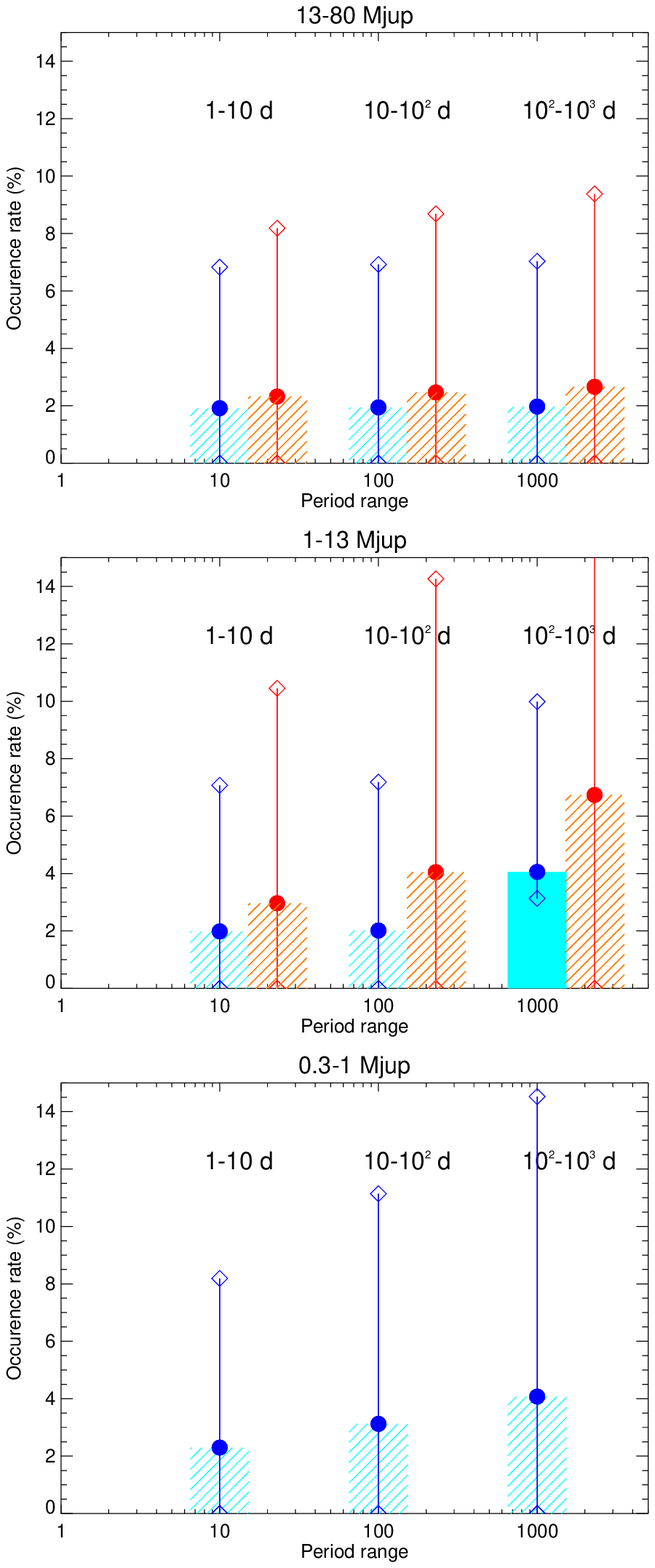}
\caption{{\it Top panel}: Companion occurrence rate in the 13 to 80-\Mjup~\msini~range for three successive (non-cumulative) $P$ ranges (from left to right: 1 to 10, 10 to 100 and 100 to 1000-day ranges). {\it Blue/cyan}: \Mstar~$\leq$ 1.5 \Msun~subsample. {\it Red/orange}: \Mstar~$\geq$ 1.5 \Msun~subsample. {\it Full color}: occurrence rate computed in the case $n_{\rm det} \geq$ 1. {\it Shaded lines}: occurrence rate computed in the case $n_{\rm det} = 0$. The $1\sigma$ confidence intervals are displayed for these two \Mstar~subsamples as blue and red solid lines, respectively. {\it Middle and bottom panels}: The same, for the 1 to 13-\Mjup~and 0.3 to 1-\Mjup~ranges, respectively.}
\label{occur_vs_P}
\end{figure}

Given that our sensitivity to GP is significantly different between our less massive (later-type) and more massive (earlier-type) targets, we decided to split our sample into two subsamples (each having roughly the same number of targets) to better investigate the impact of the stellar mass on the GP occurrence rate and properties. The first subsample is made of our 51 targets with \Mstar~$>$ 1.5 \Msun; the second subsample is made of the other 57 targets (with \Mstar~$\leq$ 1.5 \Msun). We note that the two GP systems we detected belong to the \Mstar~$\leq$ 1.5 \Msun~subsample. We hereafter discuss the results we obtained on the different ($P$,~\msini) selected domains. We sum up our main results in terms of constraints on companion occurrence rates in Table~\ref{tab_occur} and in Fig.~\ref{occur_vs_P}. Note that in the case where $n_{\rm det} = 0$, we report the companion occurrence rate with a ``$\leq$'' item in Table~\ref{tab_occur} to emphasize the difference with the $n_{\rm det} \geq 1$ case.\\

\renewcommand{\arraystretch}{1.25}
\begin{table*}[t!]
\caption{GP occurrence rate around AF dwarf stars. The parameters are displayed in normal, bold or italic fonts when considering the full star sample, the most massive (\Mstar~$>$ 1.5 \Msun) stars only or the least massive (\Mstar~$\leq$ 1.5 \Msun) stars only, respectively. For \msini~in the 0.3 to 1-\Mjup~range, we display the GP occurrence rate and other parameters only for our low-\Mstar~subsample, as our completeness is almost null in this domain for our higher-mass targets.}
\label{tab_occur}
\begin{center}
\begin{tabular}{l c c c c c l l}\\
\hline
\hline
\msini    & Orbital period  & Search       & Detected    &  Missed       & GP occurrence rate    & \multicolumn{2}{c}{Confidence intervals}\\
interval  & interval        & completeness & GP systems  & GP systems    &  (\%)                 & $1\sigma$ & $2\sigma$        \\
(\Mjup)   & (day)           &  $C$  (\%)   &             &               &                       & (\%)      &  (\%)            \\
\hline
\hline

\hline
 13-80    &  1-10           &  98        & 0 ($<$1) &  $\leq$0.02      & $\leq$1            & 0-3.8      & 0-6.9  \\
 (BD)     &                 & {\bf 96 }  & {\bf 0 ($<$1)}     &  {\bf $\leq$0.04}& {\bf $\leq$2.3}    &{\bf 0-8.2 }& {\bf 0-14.4}\\
          &                 & {\it 100}  & {\it 0 ($<$1)}     &  {\it 0}         & {\it $\leq$1.9}    &{\it 0-6.8} &{\it  0-12.1}\\
\hline
          & 1-100           &  95       &  0 ($<$1)          &  $\leq$0.05       & $\leq$1.1          & 0-4  & 0-7.2 \\
          &                 & {\bf 91}  &  {\bf 0 ($<$1)}    &  {\bf $\leq$0.1}  &{\bf $\leq$2.5}     &{\bf 0-8.6}&{\bf 0-15.2}    \\
          &                 & {\it 98}  &  {\it 0 ($<$1)}    &  {\it $\leq$0.02} &{\it $\leq$1.9}     &{\it 0-6.9}&{\it 0-12.3}     \\
\hline
          &  1-1000         &  91       &  0 ($<$1)         &  $\leq$0.1        &  $\leq$1.1         & 0-4.1     & 0-7.5 \\
          &                 & {\bf 85}  &  {\bf 0 ($<$1)}    &  {\bf $\leq$0.18} & {\bf $\leq$2.6}    &{\bf 0-9.3}&{\bf 0-16.3}    \\
          &                 & {\it 97}  &  {\it 0 ($<$1)}    &  {\it $\leq$0.03} & {\it $\leq$2}      &{\it 0-7}  &{\it 0-12.4}    \\
\hline
\hline
 1-13         &  1-10       & 86        & 0 ($<$1) &  $\leq$0.16         & $\leq$1.2        & 0-4.4      & 0-7.9 \\
(``Jupiters'')&             & {\bf 75}  & {\bf 0 ($<$1)}     &  {\bf $\leq$0.33}   & {\bf $\leq$3}    &{\bf 0-10.5}& {\bf 0-18.4}     \\
              &             & {\it 96}  & {\it 0 ($<$1)}     &  {\it $\leq$0.04}   & {\it $\leq$2}    &{\it 0-7.1} &{\it 0-12.5}     \\
\hline
          & 1-100           & 77        & 0 ($<$1)          &  $\leq$0.3          & $\leq$1.3         & 0-4.9      & 0-8.8 \\
          &                 & {\bf 57}  &  {\bf 0 ($<$1)}    &  {\bf $\leq$0.76}   &{\bf $\leq$3.9}    &{\bf 0-13.8}&{\bf 0-24.3}    \\
          &                 & {\it 95}  &  {\it 0 ($<$1)}    &  {\it $\leq$0.05}   &{\it $\leq$2}      &{\it 0-7.2} &{\it 0-12.7}    \\
\hline
          &  1-1000         & 64        & 2               &  1.1                & 3                & 2.4-7.7     & 1.3-12.6 \\
          &                 & {\bf 35}  &  {\bf 0 ($<$1)} &  {\bf $\leq$1.82}   & {\bf $\leq$6.3}  &{\bf 0-22.2} &{\bf 0-38.9}              \\
          &                 & {\it 90}   &  {\it 2}       &  {\it 0.22}         & {\it 4}          &{\it 3.1-9.9}&{\it 1.7-16}              \\
\hline
\hline
 0.3-1 (``Saturns'') &  1-10          & {\it 83}  & {\it 0 ($<$1)} &  {\it $\leq$0.2}       & {\it $\leq$2.3}  &{\it 0-8.2}&{\it 0-14.5}  \\
\hline
                     & 1-100          & {\it 63}  & {\it 0 ($<$1)} &  {\it $\leq$0.58}       &{\it $\leq$3}   &{\it 0-10.8}&{\it 0-19.1} \\
\hline
                     &  1-1000        & {\it 48}  &  {\it 0 ($<$1)}&  {\it $\leq$1.1}    & {\it $\leq$3.9}   &{\it 0-14}&{\it 0-24.9}\\
\hline
\hline
\end{tabular}
\end{center}
\end{table*}
\renewcommand{\arraystretch}{1.}

\paragraph{Companion frequency versus \msini~and versus \Mstar~--} We first look at BD companions (13 $\leq$ \msini~$\leq$ 80 \Mjup). We do not detect any BD companion in the 1 to $10^{3}$-day range, neither in our low-\Mstar~nor in our high-\Mstar~subsample. Moreover, for such \msini, our survey completeness is almost complete for the investigated periods ($C \geq$ 85\%~at any considered $P$ and \Mstar~ranges, Table~\ref{tab_occur}). This translates into a close-in BD occurrence rate of 2 to 2.6\%~(for the low and high-\Mstar~subsamples, respectively), with a 1$\sigma$ uncertainty up to 6 and 10\%, respectively. These companion frequencies are not yet very well constrained due to the relatively small number of targets per \Mstar~subsample. Still, they are in agreement with the ``brown dwarf desert'' (\ie~the paucity of BD-mass companions compared to both GP and low-mass stellar companions at sma below 3-5 au) reported around solar-like FGK dwarfs by transit and RV surveys \citep[with a $<$ 1\%~frequency, see \eg~][]{grether06}.\\

We then look at Jupiter-like and ``super-Jupiter'' companions (1 $\leq$ \msini~$\leq$ 13 \Mjup). In this \msini~domain, we find a GP occurrence rate of $3_{-0.6}^{+4.7}$\%~in the 1 to $10^{3}$-day range (corresponding roughly to sma in the 0.02 to $\sim$2.5-au range) when accounting for our whole AF sample. Yet we have to look at our two \Mstar~subsamples separately, as they are significantly different both in terms of detections and achieved sensitivity (see above). 

When looking at our low-\Mstar~subsample only, we are almost sensitive to all possible GP (completeness above 90\%); the GP occurrence rate is $4_{-0.9}^{+5.9}$\% (at a 1$\sigma$ confidence level). This value is compatible with the $\sim$$4 \pm 1$\%~GP rate found by \cite{cumming08} in roughly the same ($P$,~\msini) domain for solar-like FGK stars. Given the error bars, this result is also compatible with the GP frequency derived in more or less the same ($P$; \msini) domain and \Mstar~range by \cite{johnson10} ($\sim$7\%) or \cite{reffert14} ($\sim$4-10\%) for subgiant and giant stars, respectively. Finally, it is lower but still compatible with the GP frequency predicted by CA for this \Mstar~range \citep[moreover, the CA predictions are made considering all possible separations, see][]{kennedy08}. 

By contrast, when looking at our high-\Mstar~subsample (1.5 $\leq$ \Mstar~$\lesssim$ 3 \Msun), we do not detect any GP in the 1 to 13-\Mjup~range and at periods $\leq$ 1000 days. In this case, we still reach a completeness of about 40\%~if considering all periods (and of about 60\%~if considering only Hot Jupiters with periods below a hundred days, Table~\ref{tab_occur}). Our results translate into a GP occurrence rate of $\sim$6\%~at all periods (and of $\sim$4\%~for $P \leq 100$ days) for stars with \Mstar~between 1.5 and $\sim$3 \Msun. However, given our sample size, the uncertainties on these occurrence rates reach high values ($\sim$22 and $\sim$14\%, respectively). Therefore, our results do not allow us yet to constrain enough close-in Jovian planet occurrence rate (Fig.~\ref{occur_vs_P}) to test the proportionality relation between the stellar mass and the GP occurrence rate predicted by CA theory \citep{kennedy08} and seemingly observed for evolved stars.\\

We finally look at close-in Saturn-like GP (0.3 $\leq$ \msini~$\leq$ 1 \Mjup). Our high-\Mstar~subsample is almost insensitive to such GP (completeness $\sim$0\%). We then focus on our low-\Mstar~subsample. In this case, although our sensitivity is not as good as for more massive GP, it still reaches almost 50\%~at all periods (and it is above 60\%~for smaller periods, see Table~\ref{tab_occur}). However, we do not detect any GP in this (P,~\msini) domain. We derive a GP frequency of $3.9_{-3.9}^{+10.1}$\%~for periods below $10^{3}$ days. Therefore, we are not able to detect any significant difference with respect to the close-in Jovian GP frequency for the same \Mstar~subsample.

\paragraph{GP frequency versus orbital period --}

All of our three detected GP orbit with periods above 200 days and we do not detect any Hot Jupiter in our sample. Yet, our sensitivity is the best when considering only the shortest orbital periods ($\leq$ 10 days): for \msini~in the 1 to 13-\Mjup~range, we reach $C = 96$\%~and $C = 75$\%~for our low and high-\Mstar~subsamples, respectively. This translates into Hot Jupiter frequencies of 2 and 3\%~respectively, with uncertainties up to 7 and $\sim$10\% (Fig.~\ref{occur_vs_P}). These results are compatible with the 0.5 to $\sim$1.5\%~Hot Jupiter occurrence rate around solar-like FGK stars as derived by transit and RV surveys \citep[see \eg~][]{santerne16}. Due to the limited size of our subsamples, we cannot therefore test yet if there is a significant difference on the Hot Jupiter frequency between AF massive stars and FGK solar-like stars on the Main Sequence, as supposed by \cite{kennedy08}. As we reminded in Sect.~\ref{intro}, transiting Hot Jupiters have indeed been found around early-type MS stars. We bring here the first constraint at our knowledge on their frequency.

\section{Conclusion and perspectives}\label{conclu}

We have conducted the first systematic RV survey for brown dwarfs and giant planets with periods in the 1 to $10^{3}$-day range orbiting around early-type (AF) MS stars. We have confirmed that the challenges raised for GP detectability with RV by the early ST of these stars (smaller number of spectral lines, enhanced rotation rate) can be overcome, providing an adapted RV computation and a careful observational strategy.

Among our 108 targets, we detected and characterized one new GP system, and brought additional constraints on a second 2-GP system. We additionally detected fourteen confirmed or probable companions characterized by RV high-amplitude variations or long-term trends, and brought constraints on their minimal masses and sma. Based on these constraints, on other spectroscopic observables and on previous astrometric and/or DI detections, we concluded that most of these companions are of stellar or BD nature, though two of them can still be of planetary nature. We characterized the RV intrinsic variability of our targets, and computed the detection limits for our whole sample in the 1 to $10^{3}$-day period range. We demonstrated that close-in (sma $\lesssim$ 2.5 au) BD-mass or Jovian-mass companions are detectable around massive MS dwarfs, and that Saturn-mass companions are detectable around early dwarfs with masses up to 1.5 \Msun. We deduced the first estimation of the close-in BD and GP frequencies in the 1 to $10^{3}$-day range around AF-type stars, finding it compatible both with the ``brown dwarf desert'' and GP frequency reported in roughly the same (P,~\msini) domains for solar-like stars. However, we could not yet bring precise enough constraints to test the predictions previously made concerning the dependency of the GP occurrence rate on the stellar mass, and concerning the origin of the Hot Jupiter desert around evolved subgiant/giant stars off the Main-Sequence.\\

The next step of this study will be to associate the results (both in terms of detections and statistics) from the \sophie~northern survey to the present \harps~results. The detection of a GP on a $\sim$324-day orbit with a $\sim$2.8 \Mjup~minimal mass around the relatively young F5-6V MS dwarf \object{HD\,113337} with the \sophie~spectrograph was already reported in \cite{borgniet14}. The \sophie~performances are slightly less compared to \harps~(with a spectral resolution of $R = 75000$ and a typical RV accuracy of 5 \ms~against 1 \ms~for \harps), but they should not affect significantly a combined statistical study. Such a complete study will allow us to more than double our target sample and thus derive more significant statistical results. It should then allow us to estimate for the first time a model of the mass-period distribution of GP and BD around AF MS stars.\\

As we emphasized earlier in this paper, there is a need to test more systematically the existence of a correlation between RV-detected GP and cold debris disks indicated by IR excesses. \object{HD\,113337}, which hosts at least one GP and a debris disk \citep{chen14} is a remarkable example of such a GP+debris disk system. Some of the GP systems that we reported or discussed above (\object{HD\,60532}, \object{HD\,111998}) have not been fully investigated from this point of view so far we know. A systematic search for IR excesses in such GP systems would be useful to test the link between the presence of cold debris disks and of GP, and better understand their evolution. Younger systems with directly imaged GP at wider separations usually also host debris disks.\\

\begin{acknowledgements}
These results have made use of the SIMBAD database, operated at the CDS, Strasbourg, France. We acknowledge support from the French CNRS and the support of the Agence Nationale de la Recherche (ANR grants BLANC ANR-10-0504.01 and GIPSE ANR-14-CE33-0018). We would like to thank our co-Is on our Harps proposals: D. Ehrenreich, A. Lecavelier and G. Lo Curto. We would like to thank our anonymous referee for his judicious comments.
\end{acknowledgements}
\bibliographystyle{aa}
\bibliography{HAF_planets}

\Online

\begin{appendix}

\section{Sample}\label{whole_survey}

\onllongtab{1}{
\begin{landscape}
\begin{longtable}{p{0.7cm} p{0.7cm} c c c l | l l c c c c c c l l l l l l}
\caption{Stellar characteristics and detailed results for the 108 targets of our \harps~RV survey. Among the stellar characteristics, the spectral type (ST) and \bv~are taken from the CDS database, the \vsini~was computed with \safir~and the mass is taken from \cite{allende99}. The survey results include the observation time baseline (TBL), the number of computed spectra $N_{\rm m}$, the amplitude ({\it A}), rms and mean uncertainty $<${\it U}$>$ on the RV and BIS measurements, the mean FWHM ($<$FW$>$) and \rhk~($<$\rhk$>$), and the Pearson coefficients of the (RV,~BIS) and (RV,~\rhk) correlations. V stands for our RV variability criterion (Sect.~\ref{classif}), with V for RV variable stars and C for RV constant targets. The CL column reports the correction applied on the RV, if any ($\dagger$: correction from binary or planetary fit; $\ast$: correction from (RV,~BIS correlation); $\ast \ast$: correction from (RV,~\rhk) correlation).}\\
\hline
\multicolumn{6}{c|}{Stellar characteristics} & \multicolumn{14}{c}{Survey detailed results.}\\
\hline
HD & HIP & ST & \bv & \vsini  & Mass          & TBL & $N_{\rm m}$   & \multicolumn{3}{c}{RV}        & \multicolumn{3}{c}{BIS}      & RV-    & $<$FW$>$  & $<$\rhk$>$ & RV-   & V & CL \\

   &     &    &     &         &               &     &             & \multicolumn{3}{c}{\raisebox{.5\baselineskip}{$\overbrace{\hspace{2.7cm}}$}} & \multicolumn{3}{c}{\raisebox{.5\baselineskip}{$\overbrace{\hspace{2.7cm}}$}} & BIS & & &\rhk & & \\

   &     &    &     &         &               &     &                & {\it A} &rms  & $<${\it U}$>$ & {\it A} & rms &$<${\it U}$>$ & corr.    &             &             & corr.  &   &    \\

   &     &    &     &         &               &     &             & \multicolumn{3}{c}{\raisebox{.5\baselineskip}{$\underbrace{\hspace{2.7cm}}$}} & \multicolumn{3}{c}{\raisebox{.5\baselineskip}{$\underbrace{\hspace{2.7cm}}$}} & & & & & & \\

   &     &    &     & (\kms)  & (\Msun)       & (day)&                & \multicolumn{3}{c}{(\ms)}      & \multicolumn{3}{c}{(\ms)}     &   &  (\kms)     &  (dex)      &  &   &    \\
\hline
\hline
\endfirsthead
\caption{Continued.}\\
\hline
\multicolumn{6}{c|}{Stellar characteristics} & \multicolumn{14}{c}{Survey detailed results.}\\
\hline
HD & HIP & ST & \bv & \vsini  & Mass          & TBL & $N_{\rm m}$   & \multicolumn{3}{c}{RV}        & \multicolumn{3}{c}{BIS}      & RV-    & $<$FW$>$  & $<$\rhk$>$ & RV-   & V & CL \\

   &     &    &     &         &               &     &             & \multicolumn{3}{c}{\raisebox{.5\baselineskip}{$\overbrace{\hspace{2.7cm}}$}} & \multicolumn{3}{c}{\raisebox{.5\baselineskip}{$\overbrace{\hspace{2.7cm}}$}} & BIS & & &\rhk & & \\

   &     &    &     &         &               &     &                & {\it A} &rms  & $<${\it U}$>$ & {\it A} & rms &$<${\it U}$>$ & corr.    &             &             & corr.  &   &    \\

   &     &    &     &         &               &     &             & \multicolumn{3}{c}{\raisebox{.5\baselineskip}{$\underbrace{\hspace{2.7cm}}$}} & \multicolumn{3}{c}{\raisebox{.5\baselineskip}{$\underbrace{\hspace{2.7cm}}$}} & & & & & & \\

   &     &    &     & (\kms)  & (\Msun)       & (day)&                & \multicolumn{3}{c}{(\ms)}      & \multicolumn{3}{c}{(\ms)}     &   &  (\kms)     &  (dex)      &  &   &    \\
\hline
\hline
\endhead
\hline
\endfoot
693    & 910    & F8V      &  0.49 &  10 & 1.21 & 2124 &   33 &    16 &     4 &     1 &    11 &     3 &     3 &  0.0 &    9 & -5.27 &  0.3 & V &  \\ 
3003   & 2578   & A0V      &  0.04 & 115 & 1.87 & 2487 &   33 &   510 &   137 &    92 & 1.3 $10^{4}$ &  2949 &   229 & -0.1 &  180 &  &  & C &  \\ 
4247   & 3505   & F3V      &  0.30 &  35 & 1.64 & 2163 &   38 &    90 &    21 &    11 &   588 &   153 &    28 & -0.5 &   65 & -4.41 & -0.1 & C &  \\ 
4293   & 3521   & F0V      &  0.26 & 125 & 1.75 & 1966 &   28 &   586 &   139 &    78 &  &  &  &  &  &  &  & C &  \\ 
7439   & 5799   & F5V      &  0.46 &   8 & 1.29 & 1582 &   30 &    68 &    19 &     1 &    89 &    24 &     3 &  0.2 &   11 & -5.08 &  0.3 & V &  \\ 
9672   & 7345   & A1V      &  0.06 & 195 & 1.91 & 1779 &   24 &  1729 &   420 &   303 &  &  &  &  &  &  &  & C &  \\ 
13555  & 10306  & F5V      &  0.40 &   9 & 1.26 & 2141 &   27 &    50 &    14 &     2 &   144 &    40 &     4 & -0.2 &   13 & -5.14 &  0.4 & V &  \\ 
14943  & 11102  & A5V      &  0.20 & 115 & 1.87 & 1544 &   72 &   664 &   141 &    54 & 3.2 $10^{4}$ &  5623 &   136 &  0.0 &  156 &  &  & V &  \\ 
18978  & 14146  & A3IV-V   &  0.16 & 120 & 1.79 & 2164 &   56 &  1307 &   347 &   108 &  &  &  &  &  &  &  & V &  \\ 
19107  & 14293  & A5V      &  0.16 & 170 & 1.92 & 2163 &   28 &  2301 &   469 &   188 &  &  &  &  &  &  &  & V &  \\ 
25457  & 18859  & F7-8V    &  0.50 &  25 & 1.23 & 2143 &   45 &   174 &    46 &     3 &   163 &    43 &     7 & -0.7 &   28 & -4.39 &  0.8 & V &  \\ 
 &  &  &  &  &  &  &  &   136 &    33 &  &  &  &  &  &  &  &  0.8 & V & $\ast$ \\ 
 &  &  &  &  &  &  &  &    89 &    19 &  &  &  &  &  &  &  &  & V & $\ast\ast$ \\ 
25490  & 18907  & A0.5V    &  0.04 &  65 & 3.01 & 1991 &   38 &   401 &    96 &    57 &  &  &  &  &  &  &  & C &  \\ 
29488  & 21683  & A5V      &  0.16 & 115 & 2.08 & 1373 &   52 &   758 &   204 &    71 & 1.1 $10^{5}$ & 1.9 $10^{4}$ &   178 &  0.1 &  166 &  &  & V &  \\ 
29992  & 21861  & F3IV     &  0.37 & 100 & 1.49 & 1762 &   32 &  1146 &   379 &    27 &  3657 &   777 &    66 &  0.1 &  141 & -4.48 & -0.0 & V &  \\ 
 &  &  &  &  &  &  &  &   200 &    52 &  &  &  &  & -0.2 &  &  &  0.2 & C & $\dagger$ \\ 
30652  & 22449  & F6V      &  0.44 &  16 & 1.22 & 2163 &   55 &    73 &    17 &     3 &   137 &    28 &     7 & -0.7 &   27 & -4.76 &  0.1 & V &  \\ 
 &  &  &  &  &  &  &  &    43 &    12 &  &  &  &  &  &  &  &  0.2 & V & $\ast$ \\ 
31746  & 22844  & F5V      &  0.39 &  11 & 1.30 & 1959 &   37 &    60 &    17 &     2 &   154 &    40 &     6 & -0.8 &   19 & -4.63 & -0.1 & V &  \\ 
 &  &  &  &  &  &  &  &    40 &     9 &  &  &  &  &  &  &  &  0.0 & V & $\ast$ \\ 
32743  & 23482  & F5V      &  0.39 &  50 & 1.38 & 1988 &   36 &   116 &    37 &     5 &   350 &   103 &    13 & -0.2 &   33 & -4.76 &  0.6 & V &  \\ 
32977  & 23871  & A5IV     &  0.10 & 100 & 1.93 & 1997 &   38 &   291 &    68 &    48 & 7.5 $10^{4}$ & 1.1 $10^{4}$ &   119 &  0.1 &  150 &  &  & C &  \\ 
33256  & 23941  & F5.5V    &  0.40 &  10 & 1.26 & 1821 &   36 &    19 &     4 &     2 &    24 &     7 &     4 & -0.2 &   15 & -5.25 & -0.2 & V &  \\ 
33262  & 23693  & F9V      &  0.47 &  30 & 1.16 & 1823 &   38 &    78 &    22 &     3 &   156 &    44 &     7 & -0.8 &   24 & -4.38 &  0.4 & V &  \\ 
 &  &  &  &  &  &  &  &    44 &    13 &  &  &  &  &  &  &  &  0.4 & V & $\ast$ \\ 
38393  & 27072  & F6V      &  0.47 &   8 & 1.26 & 1546 &   39 &    19 &     4 &     1 &    35 &     7 &     4 & -0.2 &   14 & -5.11 & -0.5 & V &  \\ 
39060  & 27321  & A6V      &  0.17 & 125 & 1.75 & 4048 & 1624 &  2180 &   294 &    67 &  &  &  &  &  &  &  & V &  \\ 
40136  & 28103  & F2V      &  0.33 &  18 & 1.67 & 1395 &   33 &    38 &    10 &     3 &    90 &    20 &     7 & -0.6 &   26 & -4.47 &  0.1 & V &  \\ 
48938  & 32322  & G0V      &  0.56 &   5 & 1.15 & 1609 &   29 &     5 &     1 &     1 &    13 &     2 &     2 & -0.5 &    8 & -5.20 & -0.2 & C &  \\ 
49095  & 32366  & F6.5V    &  0.48 &   6 & 1.21 & 1951 &   50 &    49 &    16 &     1 &    21 &     5 &     3 & -0.4 &   11 & -5.33 & -0.1 & V &  \\ 
 &  &  &  &  &  &  &  &    14 &     2 &  &  &  &  & -0.3 &  &  &  0.3 & C & $\dagger$ \\ 
49933  & 32851  & F3V      &  0.35 &  12 & 1.45 & 1821 &   32 &   111 &    29 &     2 &   286 &    82 &     5 & -0.8 &   16 & -4.58 &  0.4 & V &  \\ 
 &  &  &  &  &  &  &  &    53 &    17 &  &  &  &  &  &  &  &  0.6 & V & $\ast$ \\ 
50445  & 32938  & A3V      &  0.17 &  95 & 1.70 & 1398 &   36 &   301 &    74 &    41 &  2540 &   674 &    98 &  0.1 &  150 &  &  & C &  \\ 
56537  & 35350  & A4IV     &  0.11 & 140 & 2.11 & 1824 &   39 &   609 &   175 &    76 & 1.1 $10^{5}$ & 2.2 $10^{4}$ &   189 & -0.0 &  253 &  &  & V &  \\ 
59984  & 36640  & G0V      &  0.48 &  15 & 1.16 & 1821 &   62 &    18 &     4 &     1 &    24 &     5 &     3 &  0.3 &    8 & -5.33 & -0.5 & V &  \\ 
60532  & 36795  & F6IV-V   &  0.45 &  10 & 1.50 & 1949 &  162 &   119 &    28 &     1 &    35 &     7 &     3 & -0.2 &   12 & -5.47 & -0.0 & V &  \\ 
 &  &  &  &  &  &  &  &    20 &     5 &  &  &  &  & -0.0 &  &  & -0.3 & V & $\dagger$ \\ 
60584  & 36817  & F5V      &  0.43 &  38 & 1.35 & 1821 &   39 &    79 &    24 &     9 &   324 &    87 &    23 & -0.4 &   59 & -4.65 &  0.2 & V &  \\ 
63847  & 38235  & A9V      &  0.29 &  88 & 1.54 & 1952 &   42 &  3151 &   775 &    56 & 5.3 $10^{4}$ &  7697 &   137 &  0.1 &  149 &  &  & V &  \\ 
68146  & 40035  & F6.5V    &  0.49 &   8 & 1.21 & 3023 &  194 &    19 &     4 &     1 &    43 &     7 &     3 & -0.2 &   15 & -5.22 &  0.2 & V &  \\ 
68456  & 39903  & F6V      &  0.43 &  12 & 1.31 & 1821 &   32 &  4739 &  1627 &     2 &   221 &    61 &     4 &  0.1 &   16 & -4.60 & -0.1 & V &  \\ 
 &  &  &  &  &  &  &  &   133 &    32 &  &  &  &  & -1.0 &  &  &  0.1 & V & $\dagger$ \\ 
 &  &  &  &  &  &  &  &    44 &     9 &  &  &  &  &  &  &  & -0.2 & V & $\ast$ \\ 
74591  & 42931  & A6V      &  0.19 & 115 & 1.74 & 1794 &   38 &   608 &   155 &    78 & 3.6 $10^{5}$ & 5.3 $10^{4}$ &   194 & -0.0 &  191 &  &  & C &  \\ 
76653  & 43797  & F6V      &  0.44 &  11 & 1.27 & 1823 &   42 &    54 &    13 &     2 &    75 &    16 &     5 & -0.7 &   17 & -4.55 &  0.2 & V &  \\ 
 &  &  &  &  &  &  &  &    36 &     9 &  &  &  &  &  &  &  &  0.3 & V & $\ast$ \\ 
77370  & 44143  & F4V      &  0.42 &  95 & 1.38 & 1789 &   36 &   164 &    40 &    15 &   830 &   210 &    37 & -0.2 &   85 & -4.54 &  0.3 & V &  \\ 
88955  & 50191  & A2V      &  0.05 & 105 & 2.06 & 1823 &   39 &   594 &   149 &    73 &  9817 &  2299 &   183 & -0.1 &  162 &  &  & V &  \\ 
91324  & 51523  & F9V      &  0.50 &   8 & 1.20 & 1823 &   38 &    20 &     4 &     2 &    29 &     7 &     4 & -0.4 &   15 & -5.14 &  0.0 & V &  \\ 
91889  & 51933  & F8V      &  0.52 &  31 & 1.09 & 1823 &   85 &    23 &     5 &     1 &    24 &     4 &     6 & -0.2 &    9 & -5.13 &  0.3 & V &  \\ 
93372  & 52535  & F6V      &  0.46 &  11 & 1.28 & 1959 &   47 &    35 &     8 &     2 &    58 &    12 &     5 & -0.2 &   18 & -4.78 &  0.4 & V &  \\ 
94388  & 53252  & F6V      &  0.47 &   8 & 1.22 & 1990 &   40 &    94 &    20 &     1 &   168 &    43 &     4 & -0.6 &   15 & -5.18 & -0.3 & V &  \\ 
97244  & 54688  & A5V      &  0.20 &  75 & 1.65 & 1824 &   33 &   366 &    94 &    42 &  6697 &  1515 &   101 &  0.1 &  128 &  &  & V &  \\ 
99211  & 55705  & A7V      &  0.21 & 130 & 1.78 & 1988 &   32 &   657 &   152 &    60 &  &  &  &  &  &  &  & V &  \\ 
100563 & 56445  & F5.5V    &  0.53 &  14 & 1.31 & 1720 &   27 &    32 &     9 &     2 &    67 &    18 &     5 & -0.0 &   22 & -4.85 &  0.3 & V &  \\ 
101198 & 56802  & F6.5V    &  0.52 &   5 & 1.18 & 1988 &   76 &   269 &    79 &     1 &    25 &     6 &     2 &  0.1 &    9 & -5.21 &  0.3 & V &  \\ 
 &  &  &  &  &  &  &  &    10 &     2 &  &  &  &  & -0.0 &  &  & -0.1 & V & $\dagger$ \\ 
102647 & 57632  & A3V      &  0.09 & 115 & 1.81 & 1991 &   69 &   446 &   109 &    46 &  &  &  &  &  &  &  & V &  \\ 
104731 & 58803  & F5V      &  0.41 &  20 & 1.38 & 1988 &   76 &   232 &    54 &     2 &   481 &    97 &     6 & -0.1 &   22 & -4.83 &  0.4 & V &  \\ 
105850 & 59394  & A1V      &  0.05 & 122 & 1.90 & 1888 &   35 &   833 &   227 &   114 &  &  &  &  &  &  &  & C &  \\ 
109085 & 61174  & F2V      &  0.38 &  81 & 1.49 & 1958 &   36 &   124 &    26 &    14 &   536 &   151 &    36 & -0.3 &   88 & -4.58 &  0.1 & C &  \\ 
111998 & 62875  & F6V      &  0.50 &  28 & 1.18 & 2988 &  127 &   222 &    66 &     5 &   337 &    67 &    12 &  0.1 &   42 & -4.69 &  0.1 & V &  \\ 
 &  &  &  &  &  &  &  &   112 &    18 &  &  &  &  & -0.2 &  &  &  0.2 & V & $\dagger$ \\ 
112934 & 63491  & A9V      &  0.29 &  68 & 1.49 &  748 &   20 &  3575 &   955 &    36 & 2.5 $10^{4}$ &  5702 &    90 & -0.2 &  137 & -4.33 &  0.1 & V &  \\ 
114642 & 64407  & F5.5V    &  0.46 &  13 & 1.26 & 1990 &   31 &   195 &    48 &     2 &   417 &   100 &     5 & -0.9 &   21 & -4.81 &  0.1 & V &  \\ 
 &  &  &  &  &  &  &  &    92 &    25 &  &  &  &  &  &  &  &  0.3 & V & $\ast$ \\ 
115892 & 65109  & A3V      &  0.03 &  90 & 2.01 & 1952 &   28 &   335 &    75 &    38 & 2.0 $10^{4}$ &  4409 &    96 & -0.1 &  139 &  &  & C &  \\ 
124850 & 69701  & F7III    &  0.52 &  15 & 1.18 & 3024 &  109 &   433 &    94 &     2 &   315 &    75 &     6 & -0.7 &   25 & -4.77 & -0.0 & V &  \\ 
 &  &  &  &  &  &  &  &   166 &    32 &  &  &  &  & -0.8 &  &  &  0.0 & V & $\dagger$ \\ 
 &  &  &  &  &  &  &  &   114 &    18 &  &  &  &  &  &  &  &  0.1 & V & $\ast$ \\ 
125276 & 69965  & F9V      &  0.50 &   5 & 1.18 & 2987 &   34 &    65 &    21 &     1 &    15 &     4 &     3 & -0.2 &    8 & -5.00 & -0.2 & V &  \\ 
 &  &  &  &  &  &  &  &     8 &     2 &  &  &  &  &  0.2 &  &  & -0.3 & V & $\dagger$ \\ 
128020 & 71530  & F8.5V    &  0.50 &   5 & 1.18 & 1957 &   34 &     8 &     2 &     1 &    18 &     4 &     3 & -0.2 &   11 & -5.27 &  0.3 & C &  \\ 
129685 & 72104  & A0IV     &  0.01 & 455 & 2.31 & 1776 &   21 &  8013 &  2440 &   628 &  &  &  &  &  &  &  & V &  \\ 
132052 & 73165  & F2V      &  0.32 & 105 & 1.72 & 1956 &   67 &   618 &   112 &    38 & 1.2 $10^{4}$ &  1916 &    97 & -0.0 &  166 & -4.42 &  0.0 & V &  \\ 
133469 & 73850  & F5.5V    &  0.46 &  24 & 1.24 & 1958 &   30 &    94 &    21 &     5 &   232 &    58 &    11 & -0.4 &   37 & -4.57 &  0.1 & V &  \\ 
134481 & 74376  & B9.5V    & -0.03 & 180 & 2.87 & 1776 &   20 &  3732 &   948 &   519 &  &  &  &  &  &  &  & C &  \\ 
135379 & 74824  & A3V      &  0.10 &  60 & 1.90 & 2579 &   31 &   150 &    35 &    23 &  6040 &  1634 &    58 &  0.2 &  105 &  &  & C &  \\ 
138763 & 76233  & F9IV-V   &  0.54 &   7 & 1.12 & 2608 &   46 &   201 &    56 &     2 &   218 &    53 &     5 & -1.0 &   15 & -4.37 &  0.3 & V &  \\ 
 &  &  &  &  &  &  &  &    55 &    15 &  &  &  &  &  &  &  &  0.4 & V & $\ast$ \\ 
139211 & 76716  & F6IV     &  0.48 &   7 & 1.18 & 1920 &   38 &    12 &     3 &     1 &    24 &     5 &     3 &  0.1 &   11 & -5.36 & -0.6 & V &  \\ 
141513 & 77516  & A0V      & -0.04 &  85 & 3.50 & 1785 &   25 &  1358 &   420 &    55 &   720 &   149 &   137 & -0.3 &   17 &  &  & V &  \\ 
 &  &  &  &  &  &  &  &   338 &    81 &  &  &  &  & -0.3 &  &  &  & C & $\dagger$ \\ 
142139 & 78045  & A1IV-V   &  0.06 & 110 & 2.51 & 1889 &   25 &   203 &    57 &    38 & 3.3 $10^{4}$ &  6453 &    94 &  0.0 &  131 &  &  & C &  \\ 
142630 & 78106  & B9V      &  0.06 & 215 & 1.91 & 1776 &   16 &  2146 &   585 &   464 &  &  &  &  &  &  &  & C &  \\ 
145689 & 79797  & A4IV-V   &  0.15 & 100 & 1.69 & 1889 &   26 &   898 &   218 &    66 &  &  &  &  &  &  &  & V &  \\ 
146514 & 79781  & F0IV     &  0.29 & 145 & 1.52 & 2163 &   15 &  2603 &   700 &   103 & 5.6 $10^{4}$ & 1.6 $10^{4}$ &   254 & -0.1 &  194 &  &  & V &  \\ 
146624 & 79881  & A1V      &  0.03 &  30 & 1.90 & 2609 &   38 &   111 &    24 &    17 &  &  &  &  &  &  &  & C &  \\ 
147449 & 80179  & A9III    &  0.32 &  83 & 1.67 &  389 &   36 &   321 &    71 &    20 &  2246 &   573 &    51 &  0.3 &  116 & -4.47 &  0.1 & V &  \\ 
153053 & 83187  & A5IV-V   &  0.18 &  92 & 1.74 & 1776 &   28 &   585 &   154 &    41 &  6888 &  1428 &   101 &  0.1 &  160 &  &  & V &  \\ 
153363 & 83196  & F3V      &  0.36 &  27 & 1.42 & 1995 &   85 &   885 &   221 &     8 &   787 &   188 &    20 & -0.3 &   53 & -4.58 & -0.4 & V &  \\ 
 &  &  &  &  &  &  &  &   199 &    43 &  &  &  &  & -0.5 &  &  &  0.2 & V & $\dagger$ \\ 
158352 & 85537  & A8V      &  0.21 & 165 & 1.83 & 1784 &   21 &  4763 &  1306 &    98 &  &  &  &  &  &  &  & V &  \\ 
159492 & 86305  & A5IV-V   &  0.18 &  60 & 1.73 & 2609 &   35 &   457 &    99 &    20 &  2304 &   509 &    49 & -0.0 &   82 &  &  & V &  \\ 
160613 & 86565  & A2V      &  0.09 &  95 & 2.46 & 2164 &   20 &   541 &   141 &    77 &  &  &  &  &  &  &  & C &  \\ 
164259 & 88175  & F2V      &  0.36 &  80 & 1.49 & 2609 &   34 &   190 &    63 &    18 &  1302 &   318 &    44 & -0.5 &  101 & -4.56 &  0.2 & V &  \\ 
171834 & 91237  & F3V      &  0.34 &  60 & 1.49 & 2161 &   28 &   159 &    45 &    19 &  1742 &   414 &    48 & -0.6 &  108 & -4.53 &  0.3 & V &  \\ 
172555 & 92024  & A7V      &  0.19 & 175 & 1.61 & 2618 &  110 &  1644 &   320 &    61 & 1.2 $10^{5}$ & 1.8 $10^{4}$ &   154 & -0.0 &   80 &  &  & V &  \\ 
175638 & 92946  & A5V      &  0.17 & 145 & 2.03 & 2163 &   47 &   673 &   165 &    84 &  &  &  &  &  &  &  & C &  \\ 
184985 & 96536  & F7V      &  0.45 &   5 & 1.20 & 2125 &   32 &    14 &     4 &     1 &    16 &     4 &     2 &  0.1 &   10 & -5.20 & -0.5 & V &  \\ 
186543 & 97421  & A7III-IV &  0.18 & 106 & 1.88 & 1548 &   45 &   618 &   139 &    57 &  &  &  &  &  &  &  & V &  \\ 
187532 & 97650  & F5V      &  0.38 &  95 & 1.45 & 2164 &   46 &   223 &    43 &    24 & 1.2 $10^{4}$ &  1820 &    58 &  0.1 &  109 & -4.55 &  0.0 & C &  \\ 
188228 & 98495  & A0V      & -0.06 & 115 & 2.03 & 2617 &  368 &  1936 &   291 &   138 &  &  &  &  &  &  &  & V &  \\ 
189245 & 98470  & F7V      &  0.49 & 100 & 1.22 & 2104 &   34 &   659 &   130 &    19 &  1145 &   289 &    46 & -0.0 &  107 & -4.30 & -0.1 & V &  \\ 
191862 & 99572  & F7V      &  0.48 &   8 & 1.22 & 2104 &   29 &    13 &     3 &     2 &    42 &     9 &     4 & -0.5 &   15 & -5.15 &  0.3 & C &  \\ 
192425 & 99742  & A1V      &  0.07 & 165 & 1.96 & 1775 &   20 &  1430 &   383 &   281 &  &  &  &  &  &  &  & C &  \\ 
196385 & 101808 & A9V      &  0.29 &  13 & 1.50 & 1890 &   56 &    94 &    22 &     4 &   105 &    20 &     9 & -0.3 &   21 & -4.40 & -0.2 & V &  \\ 
 &  &  &  &  &  &  &  &    45 &     9 &  &  &  &  & -0.4 &  &  & -0.2 & V & $\dagger$ \\ 
196724 & 101867 & A0V      & -0.02 &  40 & 2.57 & 1396 &    6 &  4249 &  2159 &    40 &  1711 &   696 &   101 &  0.6 &   85 &  &  & V &  \\ 
197692 & 102485 & F5V      &  0.40 &  40 & 1.34 & 1993 &   39 &   121 &    31 &     8 &   561 &   133 &    20 & -0.2 &   59 & -4.61 & -0.0 & V &  \\ 
198390 & 102805 & F5V      &  0.40 &   6 & 1.38 & 2125 &   23 &    11 &     3 &     2 &    23 &     6 &     4 & -0.1 &   14 & -5.01 &  0.0 & C &  \\ 
199254 & 103298 & A5V      &  0.11 & 145 & 2.21 & 2143 &   45 &  1816 &   419 &   110 &  &  &  &  &  &  &  & V &  \\ 
 &  &  &  &  &  &  &  &   899 &   206 &  &  &  &  &  &  &  &  & C & $\dagger$ \\ 
199260 & 103389 & F6V      &  0.50 &  13 & 1.21 & 2570 &   50 &    50 &    14 &     2 &   111 &    25 &     6 & -0.3 &   22 & -4.44 &  0.6 & V &  \\ 
202730 & 105319 & A5IV-V   &  0.18 & 210 & 1.92 & 2142 &   36 &   761 &   201 &   110 &  &  &  &  &  &  &  & C &  \\ 
203608 & 105858 & F9V      &  0.48 &   8 & 1.21 & 2002 &   38 &     7 &     2 &     1 &    17 &     4 &     3 &  0.5 &    8 & -4.99 & -0.0 & C &  \\ 
205289 & 106559 & F5V      &  0.37 &  45 & 1.38 & 2570 &   28 &   116 &    36 &    14 &   803 &   172 &    34 & -0.2 &   85 & -4.54 &  0.3 & V &  \\ 
210302 & 109422 & F6V      &  0.48 &  12 & 1.21 & 2609 &  116 &    35 &     8 &     2 &    80 &    17 &     5 & -0.2 &   22 & -5.07 & -0.2 & V &  \\ 
211976 & 110341 & F5V      &  0.42 &   5 & 1.29 & 2104 &   28 &    12 &     4 &     2 &    22 &     5 &     4 & -0.2 &   11 & -5.18 &  0.1 & V &  \\ 
213398 & 111188 & A1V      &  0.01 &  45 & 2.26 & 2446 &   22 &   131 &    35 &    21 &   419 &   137 &    53 &  0.1 &   44 &  &  & C &  \\ 
213845 & 111449 & F7V      &  0.44 &  25 & 1.35 & 2122 &   39 &   100 &    24 &     7 &   215 &    55 &    18 & -0.3 &   52 & -4.71 &  0.6 & V &  \\ 
216627 & 113136 & A3V      &  0.05 &  83 & 2.51 &  389 &   14 &  9537 &  4219 &    37 &  3329 &  1062 &    89 &  0.6 &  129 &  &  & V &  \\ 
216956 & 113368 & A4V      &  0.09 &  85 & 1.89 & 2485 &  559 &   316 &    45 &    27 &  &  &  &  &  &  &  & C &  \\ 
218396 & 114189 & F0V      &  0.26 &  49 & 1.46 & 1858 &   87 &  2597 &   816 &    35 & 6.3 $10^{4}$ &  6914 &    87 &  0.2 &   63 &  &  & V &  \\ 
219482 & 114948 & F6V      &  0.47 &   7 & 1.22 & 1965 &   31 &    74 &    24 &     2 &    75 &    20 &     4 & -0.8 &   14 & -4.41 & -0.0 & V &  \\ 
 &  &  &  &  &  &  &  &    46 &    13 &  &  &  &  &  &  &  &  0.3 & V & $\ast$ \\ 
222095 & 116602 & A2V      &  0.08 & 165 & 2.38 & 1965 &   26 &   899 &   192 &   113 &  &  &  &  &  &  &  & C &  \\ 
222368 & 116771 & F7V      &  0.50 &   7 & 1.18 & 2123 &   79 &    16 &     3 &     1 &    22 &     5 &     3 & -0.1 &   11 & -5.23 &  0.0 & V &  \\ 
223011 & 117219 & A7III-IV &  0.19 &  39 & 1.87 & 1963 &   46 &  6281 &  1729 &    14 &  1580 &   272 &    34 &  0.1 &   60 &  &  & V &  \\ 
 &  &  &  &  &  &  &  &   447 &   114 &  &  &  &  & -0.5 &  &  &  & V & $\dagger$ \\ 
224392 & 118121 & A1V      &  0.06 & 190& 1.94 & 3377 &  221 &  8118 &  1960 &   671 &  &  &  &  &  &  &  & V &  \\ 
\hline
\end{longtable}
\end{landscape}
}

\end{appendix}

\end{document}